\pdfoutput=1
\documentclass[11pt]{article}
\usepackage[margin=1in]{geometry}
\usepackage{lmodern}
\usepackage[T1]{fontenc}
\usepackage[utf8]{inputenc}
\usepackage{courier}
\usepackage{microtype}
\usepackage{xcolor}
\usepackage{amsmath,amssymb,amsthm,mathtools,amsfonts}
\usepackage{nicefrac}
\usepackage{booktabs}
\usepackage{tabularx}
\usepackage{graphicx}
\usepackage{algorithm}
\usepackage[noend]{algpseudocode}
\usepackage{subcaption}
\usepackage{enumitem}
\usepackage{aliascnt}
\usepackage{url}
\usepackage[colorlinks=true,
            linkcolor=blue!55!black,
            citecolor=blue!55!black,
            urlcolor=blue!60!black]{hyperref}
\usepackage[nameinlink,capitalize,noabbrev]{cleveref}
\AddToHook{cmd/appendix/after}{\crefalias{section}{appendix}}
\AddToHook{env/thebibliography/begin}{\raggedright}  
\hypersetup{pdftitle={Row-Private Symmetric Cone Programming: Scale-Efficient Algorithm and Lower Bound},
  pdfauthor={Deming Chu, Ruizhe Zhang}}

\newcommand{\M}{\mathcal{M}}

\newcommand{\R}{\mathbb{R}}
\newcommand{\viol}{\operatorname{viol}}

\newcommand{\PPE}{\mathsf{PrivatePerceptronEpoch}}
\newcommand{\Lap}{\mathrm{Lap}}
\newcommand{\NP}{\mathsf{NP}}
\newcommand{\BPP}{\mathsf{BPP}}
\newcommand{\Ptime}{\mathsf{P}}

\newtheorem{theorem}{Theorem}[section]
\newaliascnt{proposition}{theorem}\newtheorem{proposition}[proposition]{Proposition}\aliascntresetthe{proposition}
\newaliascnt{lemma}{theorem}\newtheorem{lemma}[lemma]{Lemma}\aliascntresetthe{lemma}
\newaliascnt{corollary}{theorem}\newtheorem{corollary}[corollary]{Corollary}\aliascntresetthe{corollary}
\newaliascnt{fact}{theorem}\newtheorem{fact}[fact]{Fact}\aliascntresetthe{fact}
\newaliascnt{assumption}{theorem}\aliascntresetthe{assumption}
\newaliascnt{definition}{theorem}\newtheorem{definition}[definition]{Definition}\aliascntresetthe{definition}
\theoremstyle{remark}
\newaliascnt{remark}{theorem}\newtheorem{remark}[remark]{Remark}\aliascntresetthe{remark}
\newtheorem*{remark*}{Remark}

\crefname{assumption}{Assumption}{Assumptions}
\Crefname{assumption}{Assumption}{Assumptions}
\crefname{fact}{Fact}{Facts}
\Crefname{fact}{Fact}{Facts}

\title{Row-Private Symmetric Cone Programming: Scale-Efficient Algorithm and Lower Bound\thanks{A Preliminary version of this paper appeared at the 40th Annual Conference on Neural Information Processing Systems (NeurIPS 2026).}}
\author{Deming Chu\thanks{Purdue University. Email: \texttt{chu292@purdue.edu}.}
  \and Ruizhe Zhang\thanks{Purdue University. Email: \texttt{rzzhang@purdue.edu}.}}
\date{}

\begin{document}

\maketitle

\begin{abstract}
Symmetric cone programming (SCP) unifies linear, second-order cone, and semidefinite programming. We study approximate feasibility for SCPs under row-level $(\varepsilon, \delta)$-differential privacy, where each private record is a linear constraint and adding or removing one row can change the feasible region discontinuously. Satisfying all constraints privately is impossible in the worst case, so we aim to find a point in the public cone that violates a small number of private constraints and satisfies the rest up to an additive slack $\eta>0$. We give a polynomial-time $(\varepsilon, \delta)$-differentially private solver for general SCPs that violates at most $\widetilde{O}\big((k^{3/2}/\varepsilon)\sqrt{\log(UR/\eta)}\big)$ constraints, where $k$ is the ambient dimension and $R, U$ bound the solution and constraint norms. This exponentially improves the $UR/\eta$ dependence in the number of violated constraints compared with prior work on covering SCPs by Song et al.\ (NeurIPS 2025). To obtain this result, we extend the private LP rescaling perceptron of Ene et al.\ (RANDOM 2025) to symmetric cones using public cone separation steps and a rank-two rescaling that stretches along a thin direction while shrinking along the average perceptron step, improving their $k^2$ dimension factor for LPs with slack to $k^{3/2}$. We also prove a lower bound of $\Omega\big((k/\varepsilon)\log\min\{UR/(k\eta),\, \varepsilon/\delta\}\big)$ violated constraints via a coordinate-wise packing argument.
\end{abstract}


\section{Introduction}
Differential privacy is the standard framework for protecting sensitive data in algorithmic computation~\cite{dwork2014algorithmic}. In many data-driven optimization problems, sensitive records directly specify the objective function or the constraints of a convex program~\cite{benvenuti2024guaranteed,benvenuti2025differentially,dvorkin2020differentially,hsu2014privately,munoz2021private}. Because the resulting solution depends on these records, releasing it directly can leak sensitive information about individual data points or constraints. This motivates the study of \emph{private convex feasibility}, where the goal is to output an (approximately) feasible point while protecting the private data that define the feasible region~\cite{ene2025solving,hsu2014privately,kaplan2024differentially}.

We study this problem for symmetric cone programs (SCPs), which unify linear programs (LPs) over nonnegative orthants, second-order cone programs (SOCPs) over Lorentz cones, and semidefinite programs (SDPs) over positive semidefinite (PSD) cones~\cite{benTalNemirovski2001,faraut1994analysis,nesterov1994interior,vandenberghe1996semidefinite}. Specifically, we focus on the \emph{row-private} setting, where the cone is public and each sensitive record specifies one linear constraint (row). A single row-private SCP solver thus covers private LP, SOCP, and SDP feasibility, addressing the question of Hsu et~al.~\cite{hsu2014privately} on whether their private LP results extend to other convex programs such as SDPs.

Two applications illustrate row-private feasibility over non-polyhedral cones. In \emph{private metric learning}~\cite{weinberger2009distance}, the decision variable is a matrix $M$ in the public PSD cone defining a Mahalanobis distance $d_M(\xi,\xi')^2=(\xi-\xi')^\top M(\xi-\xi')$. Each sensitive training triplet $(\xi,\xi^+,\xi^-)$ contributes one linear constraint on $M$,
$$
    \bigl\langle M,\ (\xi-\xi^-)(\xi-\xi^-)^\top-(\xi-\xi^+)(\xi-\xi^+)^\top\bigr\rangle \;\ge\; 1,
$$
asserting that $\xi$ is closer to $\xi^+$ than to $\xi^-$ in $d_M$. In \emph{scenario-based robust optimization}~\cite{calafiore2006scenario}, the decision variable is $x=(s,u)$ with the public Lorentz-cone constraint $\|u\|_2\le s$, and each sensitive scenario $j$ contributes one row $\langle a_j,x\rangle\le b_j$. In both examples, each private record is one row, and the public cone is non-polyhedral, so the rescaling-based row-private LP solvers~\cite{ene2025solving,kaplan2024differentially} do not apply directly.

Formally, we consider feasibility instances of the form
$$
    \langle a_i,x\rangle \le b_i \quad \forall i\in[m],
    \qquad x\in K ,
$$
where $K\subseteq\R^k$ is a $k$-dimensional symmetric cone of rank $r$ (\Cref{subsec:ejas}). The cone $K$, together with its separation oracle, is public, while the private dataset is the list of rows $(a_i,b_i)$ with $\|a_i\|_2\le U$ and $|b_i|\le U$. Privacy is measured under add/remove row adjacency, where neighboring instances differ by inserting or deleting one arbitrary bounded row. This is the high-sensitivity constraint-privacy model of Hsu et~al.~\cite{hsu2014privately}, in which one record specifies an entire halfspace. It differs from low-sensitivity perturbation models, in which the number of constraints is fixed, and a single record only slightly perturbs the right-hand sides~\cite{dvorkin2020differentially,munoz2021private}, the constraint coefficients~\cite{benvenuti2024guaranteed}, or the objective and constraints jointly~\cite{benvenuti2025differentially}. In our setting, adding or deleting one row can change the feasible region discontinuously, and can even change whether the instance is feasible.

Our notion of approximate feasibility treats the public cone and the private rows asymmetrically. Because the cone $K$ is public and encodes structural requirements such as nonnegativity, Lorentz-cone membership, or positive semidefiniteness, the output must lie exactly in $K$. The private rows, by contrast, are relaxed in two ways. First, since no private algorithm can satisfy every constraint in the worst case, the solver may violate a small number of rows. Second, each remaining row need only hold up to an additive error $\eta>0$, which we call the \emph{slack}. Allowing slack endows the relaxed system with a positive margin that measures how robustly the constraints hold simultaneously and is exploited by our algorithm. Accordingly, we measure the quality of an output $\widehat x\in K$ by its $\eta$-violation count:
$$
    \operatorname{viol}_{\eta}(\widehat x)
    := \bigl|\{i\in[m]: \langle a_i,\widehat x\rangle > b_i+\eta\}\bigr| .
$$
While differential privacy must hold for all inputs, even infeasible ones, utility is measured on instances that admit a bounded solution. Specifically, we say that an instance is \emph{$R$-feasible} if there exists a feasible point $x^\star\in K$ with $\|x^\star\|_2\le R$ (for the upper bound we take $R\ge 1$, which is without loss of generality). We call $UR/\eta$ the \emph{scale-to-slack ratio}, and say that a solver is \emph{scale-efficient} if its violation count on any $R$-feasible instance grows at most polylogarithmically in $UR/\eta$. In this work, we study how $\viol_\eta$ must depend on the dimension $k$, the privacy parameters, and the scale-to-slack ratio.

Existing polynomial-time private solvers are scale-efficient only for LPs. The private rescaling perceptrons of~Kaplan et~al.~\cite{kaplan2024differentially} and~Ene et~al.~\cite{ene2025solving} have a violation count that grows polylogarithmically with the system's inverse margin. Because relaxing the constraints by slack $\eta$ guarantees an inverse margin that is polynomial in $UR/\eta$ (\Cref{thm:margin}), their violation count is polylogarithmic in $UR/\eta$. For integer data, they can also satisfy the remaining rows exactly ($\eta=0$) by perturbing every row and privately eliminating the rows that must be tight, though this technique does not extend to general symmetric cones (\Cref{app:hardness-exact}). More fundamentally, while non-private rescaling perceptrons can handle the infinitely many supporting halfspaces of a non-polyhedral cone via a separation oracle~\cite{dadush2020rescaling,pena2016deterministic}, private LP solvers treat each constraint as a private row from a finite list of halfspaces. Conversely, the multiplicative-weights solver of~Hsu et~al.~\cite{hsu2014privately} allows a public convex set $K$ given by a linear optimization oracle, but its violation count grows polynomially in the inverse additive error. Song et~al.~\cite{song2025differential} implement this approach for covering SCPs, and when tuned to slack $\eta$, their violation count remains polynomial in the scale-to-slack ratio (\Cref{subsec:song-tuning}). We therefore ask:
\begin{center}
\emph{Is there a polynomial-time, scale-efficient row-private SCP solver?}
\end{center}

\subsection{Our results}

We answer this question affirmatively and establish two main results on row-private symmetric cone feasibility: a scale-efficient polynomial-time algorithm for general SCPs, and an information-theoretic lower bound that holds under additive slack. Our algorithm also lowers the dimension factor of private rescaling perceptrons to $k^{3/2}$ for general SCPs, which for LPs with slack improves the factor $k^2$ of Ene et~al.~\cite{ene2025solving}.

For our first result, we give, for every slack $\eta\in(0,U]$, an efficient algorithm that is $(\varepsilon,\delta)$-differentially private under add/remove row adjacency.  Its output $\widehat x$ always lies in $K$ and, on every $R$-feasible instance, satisfies with probability at least $1-\beta$
\begin{align}\label{eq:intro_upper_bound}
    \operatorname{viol}_{\eta}(\widehat x)
    \le
    \widetilde O\Bigg(
        \frac{k^{3/2}}{\varepsilon}
        \sqrt{\log\frac{UR}{\eta}}
    \Bigg)\,,
\end{align}
where $\widetilde O$ hides polylogarithmic factors in $k$, $1/\delta$, $1/\beta$, and $\log(UR/\eta)$.\footnote{For simplicity, we use $\log(UR/\eta)$ to denote $\log(2+UR\sqrt r/\eta)$, so that the bound remains meaningful when the ratio is near~$1$. And since $r\le k$, the $\sqrt r$ factor only adds a $\log k$ factor to the bound~\eqref{eq:intro_upper_bound}.} The bound does not depend on $m$, the number of constraints, and even an exponentially small slack increases it only by a polynomial factor. While its dimension factor $k^{3/2}$ (of order $n^3$ for $n\times n$ PSD matrices) is larger than that of the multiplicative-weights solver of~Song et~al.~\cite{song2025differential} for covering SCPs, our bound improves the dependence on $UR/\eta$ from polynomial to logarithmic and applies to general SCPs (\Cref{subsec:main-upper}).

Our algorithm extends the private rescaling perceptron of~Ene et~al.~\cite{ene2025solving} from LPs to SCPs. Relaxing each row by $\eta$ and homogenizing gives private homogeneous rows over the public cone $K\times\R$. The extension makes two changes to the LP algorithm. First, when an iterate lies outside the cone, the algorithm steps along a separating vector (separator) returned by the oracle of $K$, and it stops only at iterates inside the cone so that its output lies in $K$ exactly. Second, the rescaling step uses these separators together with the private rows. For the analysis, we show that moving a feasible point of norm at most $R$ slightly toward the identity element $e$ of $K$ (the identity matrix for the PSD cone) gives the homogenized system a margin of $\Omega(\eta/(UR\sqrt r))$ (\Cref{thm:margin}).

The improvement in the dimension comes from a change in the rescaling step and holds for all SCPs. For LPs with slack $\eta$, the private rescaling perceptron of~Ene et~al.~\cite{ene2025solving} violates $\widetilde O\bigl((k^2/\varepsilon)\sqrt{\log(UR/\eta)}\bigr)$ rows, so~\eqref{eq:intro_upper_bound} is smaller by a factor of $\sqrt k$. Like earlier rescaling perceptrons, the algorithm of Ene et~al.~\cite{ene2025solving} stretches space along a direction $c$ in which the feasible cone is thin. For this stretching to enlarge the volume of the cone inside the unit ball by a constant factor, the average $\mu$ of the perceptron steps in an epoch must have norm $O(1/k)$, which takes epochs of $\Theta(k^2)$ steps. Our \emph{rank-two rescaling} stretches space along $c$ and also shrinks it along $\mu$. Because every row and separator added in the epoch is nonnegative on the feasible cone and $c$ sums a subset of the terms of $\mu$, the shrinking offsets the stretching on the cone, so no point of the cone moves farther from the origin. Since the rescaling still enlarges volumes by a constant factor once $\mu$ has norm $O(1/\sqrt k)$, the volume of the cone inside the unit ball grows by a constant factor, and epochs of $\Theta(k)$ steps suffice. Moreover, we bound this volume by averaging over directions in the cone, so the noise added to $c$ enters only through its average squared projection onto these directions, which is about $k$ times smaller than its squared norm. The released direction can therefore carry $\sqrt k$ times more noise. Shorter epochs reduce the number of noisy steps from $O(k^3\log(UR/\eta))$ to $O(k^2\log(UR/\eta))$, which by advanced composition allows a $\sqrt k$ times larger privacy budget per step, and the larger noise lowers the privacy cost of each rescaling by the same factor $\sqrt k$. Both changes are needed, because the budget per step is limited both by the composition over the steps and by the composition over the rescalings, and together they give the factor $k^{3/2}$ in~\eqref{eq:intro_upper_bound}.

Our second result is an information-theoretic lower bound for the relaxed setting $\eta>0$. For solvers that satisfy all other rows with $\eta=0$, Kaplan et~al.~\cite{kaplan2024differentially} show, via a reduction to non-private PAC learning, that the violation count must grow linearly in $k$, but their argument does not apply once positive slack $\eta>0$ is allowed. We show that any $(\varepsilon,\delta)$-DP solver must incur at least
\begin{align}\label{eq:intro_lower_bound}
    \Omega\!\left(\frac{k}{\varepsilon}\,\log\min\!\left\{\frac{UR}{k\eta},\ \frac{\varepsilon}{\delta}\right\}\right)
\end{align}
$\eta$-violations on $R$-feasible LP instances, and the same bound holds up to constant factors when the public cone is a single Lorentz cone or a PSD cone of ambient dimension $k$ (which are non-polyhedral for $k\ge 3$). Whenever $\eta\le UR/(Ck)$ and $\delta\le\varepsilon/C$ for a suitable constant $C$, \eqref{eq:intro_lower_bound} is $\Omega(k/\varepsilon)$, so the linear dependence on $k$ shown by Kaplan et~al.~\cite{kaplan2024differentially} without slack ($\eta=0$) persists under additive slack. Moreover, \eqref{eq:intro_lower_bound} can be even larger than $k/\varepsilon$ when both $\eta\ll UR/k$ and $\delta\ll\varepsilon$. 

To prove this lower bound for LPs over $K=\R^k_+$, we construct hard instances in which each coordinate $x_i$ for $i\in[k]$ of the decision variable $x\in\R^k_+$ encodes a hidden scalar center $\mu_i$ chosen from $\Theta(UR/(k\eta))$ well-separated values. For each coordinate $i$, the instance repeats the pair of opposite constraints $Ux_i\le U\mu_i$ and $-Ux_i\le -U\mu_i$ a total of $t$ times. Because the $t$ copies are identical, any output $\widehat x$ with $\widehat x_i\notin[\mu_i-\eta/U,\mu_i+\eta/U]$ violates all $t$ copies of one of the two constraints, incurring $t$ violations on coordinate $i$. Conversely, if $\widehat x$ incurs fewer than $kt/2$ violations in total, then $\widehat x_i$ must lie in $[\mu_i-\eta/U,\mu_i+\eta/U]$ for at least half of the coordinates, uniquely revealing those centers $\mu_i$ and contradicting group privacy over the $2t$ rows per coordinate. For a single Lorentz or PSD cone, we obtain the same bound by fixing an interior point of $K$ and imposing these opposite constraints along $k-1$ orthogonal directions around it (\Cref{subsec:main-lower}).

Finally, our upper bound grows as $k^{3/2}$, whereas the lower bound is linear in $k$. A computationally unbounded algorithm, the exponential mechanism, matches the $k/\varepsilon$ factor of the lower bound (\Cref{rmk:lower-tight}), so the gap concerns polynomial-time algorithms and closing it is the main open question. We note that proving a stronger lower bound for efficient algorithms would require breakthroughs in complexity theory. Indeed, for LPs with $\varepsilon=1$ and scale-to-slack ratio polynomial in $k$, an unconditional lower bound of $k^{1+\Omega(1)}$ violations for polynomial-time algorithms would imply $\NP\not\subseteq\BPP$ and hence $\Ptime\ne\NP$ (\Cref{rmk:np-barrier}).

\section{Related Work}
\label{sec:related-work}

\paragraph{Private linear programming.}
Hsu et~al.~\cite{hsu2014privately} initiated the systematic study of private LPs, with separate models in which the constraints, the right-hand sides, or the
objective are private. A subsequent line of work~\cite{benvenuti2024guaranteed,benvenuti2025differentially,dvorkin2020differentially,munoz2021private} studies low-sensitivity models and gives private solutions with feasibility guarantees. Closer to our setting, Kaplan et~al.~\cite{kaplan2024differentially} and~Ene et~al.~\cite{ene2025solving} privatize rescaling perceptrons~\cite{dunagan2006simple,hoberg2017improved} and obtain efficient row-private LP algorithms that violate a small number of rows and, for integer data, satisfy all other rows without slack ($\eta=0$). For packing and covering LPs, Ene et~al.~\cite{ene2026solving} use dense multiplicative weights to obtain a solver that violates $\widetilde O(k^{3/2}/(\varepsilon\eta^{3.5}))$ rows while satisfying the remaining rows up to a factor of $1\pm\eta$.

\begin{table}[t]
\centering
\small
\setlength{\tabcolsep}{4.5pt}
\renewcommand{\arraystretch}{1.15}
\begin{tabular*}{\textwidth}{@{\extracolsep{\fill}} l l l c c @{}}
\toprule
 & & & \multicolumn{2}{c}{\textbf{Violated rows}} \\
\cmidrule(lr){4-5}
\textbf{Ref.} & \textbf{Problem setting} & \textbf{Non-violated rows} & \textbf{Upper bound} & \textbf{Lower bound} \\
\midrule
\cite{hsu2014privately}         & LP over public convex $K$            & Additive slack $\eta$         & $\widetilde{O}(\mathrm{OPT}^2 / (\varepsilon\eta^2))$      & --- \\
\addlinespace[2pt]
\cite{kaplan2024differentially} & LP feasibility                       & Exact (integer data)          & $\widetilde{O}(k^9 / \varepsilon)$                         & $\Omega(k)$ \\
\addlinespace[2pt]
\cite{ene2025solving}           & LP feasibility                       & Exact (integer data)          & $\widetilde{O}(k^4 / \varepsilon)$                         & --- \\
                                &                                      & Additive slack $\eta$         & $\widetilde{O}(k^2 / \varepsilon)$                         & --- \\
\addlinespace[2pt]
\cite{song2025differential}     & Covering SCP & Additive slack $\eta$         & $\widetilde{O}(k_C\,\mathrm{OPT}^2 / (\varepsilon\eta^2))$ & --- \\
\addlinespace[2pt]
\cite{ene2026solving}           & Packing / covering LP                & Multiplicative ($1 \pm \eta$) & $\widetilde{O}(k^{3/2} / (\varepsilon\eta^{3.5}))$         & --- \\
\midrule
\textbf{This work}              & SCP feasibility              & Additive slack $\eta$         & $\boldsymbol{\widetilde{O}(k^{3/2} / \varepsilon)}$                     & $\boldsymbol{\Omega(k / \varepsilon)}$ \\
\bottomrule
\end{tabular*}
\caption{Comparison with prior work on $(\varepsilon,\delta)$-row-private LP and SCP solvers. Here, $k$ is the ambient dimension (the number of variables in LPs), and $\mathrm{OPT}$ is the optimal value of a packing or covering LP/SCP with $b_i=1$. While~\cite{hsu2014privately} gives an oracle-based framework for LPs over a public convex set $K$, an explicit bound is derived only for unit-cost fractional set cover ($\mathrm{OPT}$ public). For~\cite{song2025differential}, $k_C = r$ for PSD cones and $k-1$ for Lorentz cones. In~\cite{ene2026solving}, non-violated rows satisfy $a_i^\top x \le (1+\eta)b_i$ for packing and $a_i^\top x \ge (1-\eta)b_i$ for covering.} 
\label{tab:comparison}
\end{table}

\paragraph{Private symmetric cone programming.}
Song et al.~\cite{song2025differential} give the first general framework for differentially private SCPs, based on Gaussian mechanisms over Euclidean Jordan algebras and multiplicative weights. In the row-private model, they prove guarantees only for covering SCPs over a simple (irreducible) Jordan algebra, and~\Cref{subsec:main-upper} compares their bound with ours.

\paragraph{Rescaling perceptrons and conic feasibility.}
The classical perceptron needs a number of iterations polynomial in the inverse margin. Rescaling perceptrons~\cite{dunagan2006simple,hoberg2017improved,pena2016deterministic} reduce this dependence to logarithmic by repeatedly finding a direction in which the feasible region is thin and stretching space along it. Hoberg and Rothvoss~\cite{hoberg2017improved} also stretch space along several directions at once, and for systems in $\R^d$ they rescale once a convex combination of the rows has norm $O(1/d)$. Reaching this threshold takes $\Theta(d^2)$ perceptron steps, or $\widetilde O(d)$ iterations of accelerated methods that use all rows in every iteration~\cite{dadush2020rescaling,hoberg2017improved}. Belloni et~al.~\cite{belloni2009efficient} give a rescaling perceptron for conic systems $Ax\in\operatorname{int}K$ that is efficient given a deep-separation oracle, constructing such oracles for products of halfspaces and second-order cones and a randomized one for semidefinite cones. Pe{\~n}a and Soheili~\cite[Section~3]{pena2016deterministic} show that, for general convex cones, an ordinary separation oracle suffices when the algorithm rescales along a sum of separators collected by the perceptron. Dadush et~al.~\cite{dadush2020rescaling} also rescale with collected separators in an oracle model, and related projection-and-rescaling algorithms have been developed for symmetric cones~\cite{lourenco2019extension,pena2017solving}. Our algorithm incorporates this use of separators into the private rescaling perceptron of~Ene et~al.~\cite{ene2025solving}, so that our rescaling step combines separators with noisy private rows. Unlike these rescalings, our rank-two rescaling also shrinks space along the average perceptron step, so this average needs norm only $O(1/\sqrt d)$, which $\Theta(d)$ noisy perceptron steps reach.

\paragraph{Lower bounds.}
Kaplan et~al.~\cite{kaplan2024differentially} show that a row-private LP solver must violate a number of rows linear in $k$. Their proof combines a black-box reduction to non-private PAC learning of halfspaces with the $\Omega(k/m)$ error-rate bound from the fundamental theorem of statistical learning~\cite{ehrenfeucht1989general,shalev2014understanding}. The reduction identifies a violated row with a misclassified example, so it applies only when all other rows hold without slack ($\eta=0$). Fingerprinting codes~\cite{bun2014fingerprinting,dwork2015robust}, including the padding-and-permuting codes of Peter et~al.~\cite{peter2024smooth}, are a standard tool for lower bounds on the error of $(\varepsilon,\delta)$-DP algorithms. Our lower bound (\Cref{subsec:main-lower}) instead uses a coordinate-wise packing argument in the spirit of Hardt and Talwar~\cite{hardt2010geometry} and Acharya et~al.~\cite{acharya2021differentially}, in which $\delta$ enters only through group privacy and yields the $\log(\varepsilon/\delta)$ term.

\section{Preliminaries}

This section reviews symmetric cones, Euclidean Jordan algebras, and separation oracles (\Cref{subsec:ejas}), recalls differential privacy (\Cref{subsec:dp}), and formally defines the row-private SCP feasibility problem (\Cref{subsec:conic-feasibility}).

\subsection{Symmetric cones and Euclidean Jordan algebras}
\label{subsec:ejas}

\paragraph{Symmetric cones and Jordan structure.}
For a closed convex cone $C\subseteq\R^d$, the \emph{dual cone} is
$
C^*:=\{h\in\R^d:\langle h,z\rangle\ge 0\text{ for all }z\in C\}.
$
A closed convex cone $K\subseteq\R^k$ with nonempty interior is \emph{symmetric} if it is \emph{self-dual} ($K^*=K$) and \emph{homogeneous} (any interior point of $K$ can be mapped to any other by an invertible linear map that preserves $K$)~\cite{faraut1994analysis}.  \Cref{tab:cones} lists the three standard examples, and products of symmetric cones are again symmetric.

Every symmetric cone is the cone of squares $K=\{x\circ x:x\in J\}$ of a \emph{Euclidean Jordan algebra} (EJA) $J$, a real inner-product space of dimension $k$ with a bilinear product $\circ$~\cite{faraut1994analysis}.  For the PSD cone, $J=\mathbb S^n$ with $X\circ Y=(XY+YX)/2$.  As for symmetric matrices, every $x\in J$ has $r$ real eigenvalues, counted with multiplicity, where $r$ is the \emph{rank} of $J$ and of $K$ ($r=n$ for $\R_+^n$ and for the PSD cone of $n\times n$ matrices, and $r=2$ for every Lorentz cone).  We write $\lambda_{\min}(x)$ for the smallest eigenvalue of $x$.  Rank-one projectors $vv^\top$, orthonormal eigenbases, and the identity matrix generalize to the \emph{primitive idempotents}, \emph{Jordan frames}, and the \emph{identity element} $e$ of $J$.  We equip $J$ with its canonical trace inner product (\Cref{subsec:eja-background}), which defines all inner products and norms (including the bounds $U$ and $R$ below) throughout the paper.  Our margin bound (\Cref{thm:margin}) uses four standard properties:
\begin{enumerate}
\item[(J1)] \emph{Self-duality:} $K=K^*$.
\item[(J2)] \emph{Spectral decomposition:} every \(x\in J\) can be written as \(x=\sum_{j=1}^r\lambda_j q_j\), where \(\lambda_1,\ldots,\lambda_r\in\R\) are the eigenvalues of \(x\) and \(q_1,\ldots,q_r\in K\) form a \emph{Jordan frame}, meaning that they are mutually orthogonal primitive idempotents of unit norm with \(\sum_j q_j=e\). Hence \(x-\lambda_{\min}(x)\,e=\sum_j(\lambda_j-\lambda_{\min}(x))\,q_j\in K\).
\item[(J3)] \emph{Variational form of $\lambda_{\min}$:}
      $\lambda_{\min}(x)=\min_{q\in Q}\langle q,x\rangle$, where $Q\subseteq K$ is
      the set of primitive idempotents.  Hence $\lambda_{\min}$ is concave, and, by (J1) and (J2),
      $x\in K$ if and only if $\lambda_{\min}(x)\ge 0$.
\item[(J4)] \emph{Identity element:} $\|e\|_2=\sqrt r$ and $\lambda_{\min}(e)=1$.  If $u\in K$
      has unit norm, then (J2) and (J3) give $u=\sum_j\lambda_jq_j$ with $\lambda_j\ge 0$ and $\sum_j\lambda_j^2=1$,
      so $\langle e,u\rangle=\sum_j\lambda_j\ge 1$.
\end{enumerate}

\begin{table}[ht]
\centering
\caption{Three standard symmetric cones, with ambient dimension $k$ and rank $r$.  Here $\mathbb S^n$ is the space of real symmetric $n\times n$ matrices.  
}
\label{tab:cones}
\begin{tabular}{@{}llcc@{}}
\toprule
Cone & Definition & Ambient dim.\ $k$ & Rank $r$ \\
\midrule
Nonnegative orthant & $\R_+^n$ & $n$ & $n$ \\
Lorentz cone        & $\{(s,u)\in\R\times\R^{n-1}:s\ge\|u\|_2\}$ & $n$ & $2$ \\
PSD cone            & $\{X\in\mathbb S^n:X\succeq 0\}$ & $n(n+1)/2$ & $n$ \\
\bottomrule
\end{tabular}
\end{table}

\paragraph{Separation oracles.}
A \emph{separation oracle} for a closed convex cone \(C\), on input \(x\), either reports \(x\in C\) or returns a unit vector \(h\in C^*\) with \(\langle h,x\rangle<0\), which we call a \emph{unit separator} (or simply a \emph{separator}). Such an \(h\) exists for every \(x\notin C\) by the bipolar theorem \(C^{**}=C\). For a general symmetric cone, (J1) and (J2) give a separation oracle: if \(\lambda_{\min}(x)<0\), any frame element \(q_j\) of (J2) with \(\lambda_j=\lambda_{\min}(x)\) lies in \(K=K^*\) and satisfies \(\langle q_j,x\rangle=\lambda_j<0\). The three standard cones admit direct oracles that avoid a full spectral decomposition. For the PSD cone, a pivoted \(LDL^\top\) factorization decides whether \(X\succeq 0\) in \(O(n^\omega)\) arithmetic operations ($\omega= 2.25$ is the matrix multiplication exponent~\cite{openai2026matmul,openai2026matmullb}) and otherwise yields a vector \(v\) with \(v^\top Xv<0\), from which \(h:=vv^\top/\|v\|_2^2\) is a unit separator. For the orthant and the Lorentz cone, membership and separation take \(O(n)\) operations. For a product cone \(K=K_1\times\cdots\times K_p\), a block \(x_j\notin K_j\) has a separator \(h_j\in K_j^*\), and padding it with zeros gives a separator in \(K^*\).

\subsection{Differential privacy}
\label{subsec:dp}

\begin{definition}[Differential privacy~\cite{dwork2014algorithmic}]
Two lists of rows are \emph{neighbors} if one is obtained from the other by adding or removing one row.  A randomized algorithm $\M$ is \emph{$(\varepsilon,\delta)$-differentially private} (DP) if for all neighbors $D,D'$ and every event $S$,
$$
\Pr[\M(D)\in S]\le e^\varepsilon\Pr[\M(D')\in S]+\delta .
$$
The case $\delta=0$ is called \emph{pure} differential privacy.
\end{definition}

\paragraph{Sensitivity and noise mechanisms.}
We release statistics of the private rows by adding noise calibrated to their \emph{sensitivity}, the largest change in the statistic when one row is added or removed (\Cref{subsec:privacy-tools}).  For example, the number of rows violated at a fixed point has $\ell_1$-sensitivity $1$, and the Laplace mechanism (\Cref{fact:laplace-mechanism}) releases it under $(\varepsilon_1,0)$-DP with noise of scale $1/\varepsilon_1$.  The average of $N$ unit-norm rows has $\ell_2$-sensitivity at most $2/N$.  On inputs with $N\ge\nu$ for a public threshold $\nu$, this is at most $2/\nu$, and the Gaussian mechanism (\Cref{fact:gaussian-mechanism}) releases the average under $(\varepsilon_1,\delta_1)$-DP with independent Gaussian noise of standard deviation $O(\sqrt{\log(1/\delta_1)}/(\nu\varepsilon_1))$ per coordinate.  The proof of \Cref{prop:privacy-composition} makes this precise by dividing by $\max\{N,\nu\}$ instead of $N$.

\paragraph{Composition, post-processing, and group privacy.}
By advanced composition (\Cref{fact:advanced-composition}), for $\varepsilon\le 1$ and $\delta'\in(0,1/2]$, $L$ adaptively chosen steps that are each $(\varepsilon_1,\delta_1)$-DP with $\varepsilon_1\le\varepsilon/\sqrt{8L\log(1/\delta')}$ are together $(\varepsilon,\,L\delta_1+\delta')$-DP.  The per-step privacy budget thus scales as $1/\sqrt L$ rather than $1/L$.  We also rely on two standard properties~\cite{dwork2014algorithmic}: post-processing, which covers public separation-oracle steps in our algorithm, and group privacy, which underlies our lower bound.

\begin{fact}[Post-processing]
\label{fact:post-processing}
If $\M$ is $(\varepsilon,\delta)$-DP and $F$ is a randomized map that does not access the private data, then $F\circ\M$ is $(\varepsilon,\delta)$-DP.
\end{fact}

\begin{fact}[Group privacy]
\leavevmode\phantomsection\label{fact:group-privacy}%
If $\M$ is $(\varepsilon,\delta)$-DP and $D=D_0,D_1,\ldots,D_L=D'$ are lists in which $D_{i-1}$ and $D_i$ are neighbors for every $i\in[L]$, then for every event $S$,
$\Pr[\M(D)\in S]\le e^{L\varepsilon}\Pr[\M(D')\in S]+\delta\sum_{i=0}^{L-1}e^{i\varepsilon}$.
\end{fact}

\subsection{Row-private symmetric cone feasibility}
\label{subsec:conic-feasibility}

\begin{definition}[Row-private SCP feasibility]
\label{def:problem}
An instance consists of a public symmetric cone $K\subseteq\R^k$ of rank $r$, accessed through a separation oracle, and a private list of rows $(a_i,b_i)\in\R^k\times\R$, $i\in[m]$, with $\|a_i\|_2,|b_i|\le U$.  The instance is \emph{$R$-feasible} if some witness $x^\star\in K$ with $\|x^\star\|_2\le R$ satisfies $\langle a_i,x^\star\rangle\le b_i$ for all $i\in[m]$ ($R$ bounds the norm of $x^\star$, not of the released point).  For a slack $\eta>0$, row $i$ is \emph{$\eta$-satisfied} at $\widehat x\in K$ if $\langle a_i,\widehat x\rangle\le b_i+\eta$ and \emph{$\eta$-violated} otherwise, and the \emph{$\eta$-violation count} is
$$
\viol_\eta(\widehat x):=\bigl|\{i\in[m]:\langle a_i,\widehat x\rangle>b_i+\eta\}\bigr|.
$$
Given public $U,R,\eta>0$ and privacy parameters $\varepsilon,\delta$, the goal is an $(\varepsilon,\delta)$-DP algorithm that outputs a point $\widehat x\in K$ and, on every $R$-feasible instance, achieves small $\viol_\eta(\widehat x)$ with high probability.
\end{definition}

Our algorithm (\Cref{app:mixed-solver}) accesses $K$ only through its separation oracle, and its analysis uses the EJA structure of $K$ only in the margin bound (\Cref{thm:margin}).  Both of our bounds are stated in terms of the ambient dimension $k$, which for the PSD cone is quadratic in the rank $r=n$.  In the upper bound, the rank appears only inside the logarithmic factor \(\sqrt{\log(UR\sqrt r/\eta)}\).

\section{Main Results and Technical Overview}
\label{sec:main-results}

This section states our upper and lower bounds and sketches their proofs.  The formal statements and proofs are in \Cref{app:mixed-solver,app:upper-proof,app:lower-bound,app:np-barrier}, and \Cref{app:hardness-exact} explains why the LP method for zero slack (\(\eta=0\)) does not extend to general SCPs.

\subsection{Upper bound}
\label{subsec:main-upper}

\begin{theorem}[Upper bound, informal version of \Cref{thm:main}]
\label{thm:upper-informal}
Let $R\ge 1$, $\eta\in(0,U]$, and $\varepsilon,\delta,\beta\in(0,1)$. There is an efficient $(\varepsilon,\delta)$-DP algorithm under add/remove row adjacency that always outputs a point $\widehat x\in K$ and, on every $R$-feasible instance of \Cref{def:problem}, satisfies with probability at least $1-\beta$
\begin{align*}
\viol_\eta(\widehat x)
&\le \widetilde O\!\left(\frac{k^{3/2}}{\varepsilon}\sqrt{\log\frac{UR\sqrt{r}}{\eta}}\right),
\end{align*}
where $\widetilde O(\cdot)$ hides factors polylogarithmic in $k$, $1/\delta$, $1/\beta$, and $\log(UR\sqrt{r}/\eta)$.
\end{theorem}

The algorithm makes $O\bigl(k^2\log(UR\sqrt{r}/\eta)\bigr)$ separation-oracle queries, each answered by a single call to the separation oracle for $K$, as shown in \Cref{subsec:sep-oracle}. Following~\cite{kaplan2024differentially}, we measure running time in the real-arithmetic model with exact oracle calls and continuous Gaussian and Laplace samples, in which the runtime is polynomial in $m$, $k$, $1/\varepsilon$, $\log(1/\delta)$, $\log(UR\sqrt{r}/\eta)$, and $\log(1/\beta)$. For small $\delta$, the $\widetilde O(\cdot)$ notation hides $\log^2(1/\delta)$, arising from $\sqrt{\log(1/\delta)}$ in advanced composition and $\log^{3/2}(1/\delta)$ in the stopping threshold.

\begin{remark}[The role of $\eta$]
\label{rmk:upper-slack}
\Cref{thm:upper-informal} relaxes the private rows in two ways: at most $\viol_\eta(\widehat x)$ rows may be violated arbitrarily, while all remaining rows hold up to additive slack $\eta$. The user-chosen parameter $\eta$ enters the bound only through $\sqrt{\log(UR\sqrt{r}/\eta)}$, so even an exponentially small slack $\eta=2^{-L}$ increases the violation count only by $\mathrm{poly}(L)$. Moreover, $\eta>0$ is used solely to guarantee a positive conic margin: whenever a positive lower bound on the homogenized margin at $\eta=0$ is known in advance, the same algorithm satisfies all non-violated rows exactly. Whether $\eta>0$ can be avoided on general non-polyhedral SCPs with zero margin remains open; see \Cref{sec:limitations}.
\end{remark}
\subsubsection{Our techniques for the upper bound}
We now sketch the proof of \Cref{thm:upper-informal}, with full proofs deferred to \Cref{app:mixed-solver,app:upper-proof}. The argument proceeds in three steps: reducing relaxed SCP feasibility to a homogeneous system with positive mixed margin, identifying the two $k^2$ bottlenecks in the LP rescaling perceptron, and removing both via cone separation and rank-two rescaling (\Cref{alg:private-perceptron-informal}).

\paragraph{Step 1: Homogenization and positive mixed margin.}
Writing $y=(x,t)\in J\times\R$, we replace each relaxed constraint $\langle a_i,x\rangle\le b_i+\eta$ with the homogeneous inequality $\langle\bar a_i,y\rangle>0$, where $\bar a_i$ normalizes $(-a_i,\,b_i+\eta)$ to unit length, and work over the public cone $C_K:=K\times\R$ of dimension $d=k+1$. Whenever $t>0$, having $\langle\bar a_i,y\rangle>0$ implies $\langle a_i,x/t\rangle<b_i+\eta$. Appending copies of the public row $e_t=(0,1)$ beyond the algorithm's violation budget enforces $t>0$ with high probability, allowing us to output $\widehat x:=x/t$, or $0\in K$ if $t\le 0$ or the run aborts; see \Cref{app:upper-proof}. Convergence of the homogenized solver is governed by the following condition measure, studied in the non-private conic literature by Belloni et~al.~\cite{belloni2009efficient}.

\begin{definition}[Mixed margin]
\label{def:mixed-margin}
Let $\bar A\in\R^{m\times d}$ have unit rows $\bar a_1,\ldots,\bar a_m$, and let $C\subseteq\R^d$ be a closed convex cone. The \emph{mixed margin} of $(\bar A,C)$ is
\begin{align*}
\rho(\bar A,C)
&:= \max_{y\in C,\ \|y\|_2\le 1}
\min\Bigl\{
\min_{i\in[m]}\langle \bar a_i,y\rangle,\
\inf_{h\in C^*,\ \|h\|_2=1}\langle h,y\rangle
\Bigr\},
\end{align*}
with the convention $\inf_\varnothing=+\infty$, so that when $C=\R^d$ and $C^*=\{0\}$, the mixed margin reduces to the standard linear margin of $\bar A$.
\end{definition}

Geometrically, when $\rho(\bar A,C)>0$, it equals the width of the feasible cone $F:=\{y\in C:\bar Ay\ge 0\}$, i.e., the radius of the largest Euclidean ball contained in $F$ with center in the unit ball. For LPs, the orthant constraints $x\ge 0$ comprise $k$ halfspaces that can be absorbed directly into $\bar A$, recovering the classical LP margin~\cite{dunagan2006simple,hoberg2017improved,pena2016deterministic}; for non-polyhedral cones, the dual-cone infimum replaces these facet rows.

In \Cref{thm:margin}, we prove that any $R$-feasible instance yields mixed margin $\rho=\Omega\bigl(\eta/(UR\sqrt{r})\bigr)$. Whereas $\eta=0$ on integer LPs requires vertex and Cramer's-rule arguments~\cite{ene2025solving}, positive slack lets us construct an interior witness directly by nudging a feasible point $x^\star\in K$ toward the scaled Jordan identity $(R/\sqrt{r})\,e$ from \Cref{subsec:ejas}: let
\begin{align*}
x_\eta := (1-\alpha)\,x^\star + \alpha\,\frac{R}{\sqrt{r}}\,e,
\qquad
\alpha := \frac{\eta}{2U(R+1)}.
\end{align*}
Because $x^\star$ is feasible, $|b_i|\le U$, and $|\langle a_i,(R/\sqrt{r})e\rangle|\le UR$, the point $x_\eta$ violates each unrelaxed row by at most $\alpha U(R+1)=\eta/2$, leaving $\eta/2$ of slack after $\eta$-relaxation. Meanwhile, concavity of $\lambda_{\min}$ in (J3) and $\lambda_{\min}(e)=1$ in (J4) imply $\lambda_{\min}(x_\eta)\ge\alpha R/\sqrt{r}$, so for every unit normal $u\in K^*=K$ from (J1), properties (J2) and (J4) yield $\langle u,x_\eta\rangle\ge\lambda_{\min}(x_\eta)\langle u,e\rangle\ge\lambda_{\min}(x_\eta)$. Since $(-a_i,b_i+\eta)$ has norm $O(U)$ and $(x_\eta,1)$ has norm $O(R)$ for $R\ge 1$ and $\eta\le U$, normalizing gives $\rho=\Omega\bigl(\eta/(UR\sqrt{r})\bigr)$. Half of the explicit bound in \Cref{thm:margin} depends only on public parameters and serves as the public margin input $\rho_0$.

\paragraph{Step 2: Private rescaling perceptron and the $k^2$ bottleneck.}
To motivate our solver, recall how the rescaling perceptron solves a homogeneous LP $\bar Ay>0$ of margin $\rho$ in dimension $d$~\cite{dunagan2006simple,hoberg2017improved,pena2016deterministic}. Standard perceptron updates from $y=0$ require up to $\Theta(1/\rho^2)$ steps. Instead, the rescaling perceptron runs in epochs of $T=\Theta(d^2)$ steps: if an epoch fails to find a feasible point, the added unit rows nearly cancel, so their average $\mu$ satisfies $\|\mu\|_2=O(1/\sqrt{T})=O(1/d)$. Because $F$ lies in the nonnegative halfspace of every added row, $F$ must be thin along a sub-sum $c$ of the terms of $\mu$ lying on one side of a random hyperplane, which satisfies $0\le\langle c,z\rangle\le\langle\mu,z\rangle$ on $F$ and $\|c\|_2=\Omega(1/\sqrt{d})$ with constant probability. Stretching space along $c$ enlarges $\mathrm{vol}(F\cap\mathbb B^d)$ by a constant factor provided $\langle c,z\rangle=O(1/d)$ for all unit $z\in F$, which follows from $\|\mu\|_2=O(1/d)$ and forces $T=\Theta(d^2)$. By \Cref{lem:initial-volume}, $\mathrm{vol}(F\cap\mathbb B^d)$ starts at least $(\rho/2)^d\,\mathrm{vol}(\mathbb B^d)$ and never exceeds $\mathrm{vol}(\mathbb B^d)$, bounding the number of epochs by $O\bigl(d\log(1/\rho)\bigr)$.

In the private adaptation of~\cite{ene2025solving}, each step checks via a noisy count whether more than $\nu$ rows are violated—halting with $O(\nu)$ violations otherwise—and moves along the average of the violated rows plus Gaussian noise. At per-step privacy budget $\varepsilon_1$, keeping the Gaussian noise norm $O\bigl(\sqrt{d}/(\nu\varepsilon_1)\bigr)$ below the unit signal requires $\nu\approx\sqrt{d}/\varepsilon_1$. Two separate bottlenecks then force $\varepsilon_1\approx\varepsilon/\bigl(d^{3/2}\sqrt{\log(1/\rho)}\bigr)$ up to logarithmic factors:
\begin{enumerate}[label=(\alph*)]
\item composing $O\bigl(d\log(1/\rho)\bigr)$ epochs of $T=\Theta(d^2)$ private steps totals $O\bigl(d^3\log(1/\rho)\bigr)$ steps; and
\item keeping the Euclidean noise norm on the rescaling vector within $O(1/d)$ requires spending privacy budget $d\varepsilon_1$ on each of the $\widetilde O\bigl(d\log(1/\rho)\bigr)$ rescaling trials.
\end{enumerate}
Both constraints yield $\nu=\widetilde O\bigl((d^2/\varepsilon)\sqrt{\log(1/\rho)}\bigr)$.

\begin{algorithm}[t]
\caption{Private rescaling perceptron for SCP feasibility (informal).}
\label{alg:private-perceptron-informal}
\begin{algorithmic}[1]
\For{$O(d\log(1/\rho_0))$ epochs} \Comment{$\rho_0$: public margin bound}
    \State $y\gets 0$
    \For{$T=\Theta(d)$ steps}
        \If{$y\notin C_K$} \Comment{public branch}
            \State $g\gets p$, the unit separator returned by the oracle at $y$
        \Else
            \State \textbf{return} $y$ if the noisy violation count at $y$ is at most $\nu$ \Comment{undo rescaling; output $x/t$}
            \State $g\gets$ average of the rows violated by $y$
        \EndIf
        \State $y\gets y+g+\zeta$, \ $\zeta\sim\mathcal N(0,\sigma^2I)$
    \EndFor
    \State $\widehat\mu\gets y/T$ \Comment{$\widehat\mu\approx\sum_j\widetilde\lambda_j\widetilde a_j$}
    \State $\widehat c\gets\sum_{j\in S}\widetilde\lambda_j\widetilde a_j+\gamma$ for a random index set $S$ and $\gamma\sim\mathcal N(0,\theta^2I)$
    \State Rescale space by $M:=\bigl(I+(\widehat c\,\widehat c^\top-4\widehat\mu\widehat\mu^\top)/\|\widehat c\|_2^2\bigr)^{1/2}$ \Comment{rank-two rescaling}
\EndFor
\end{algorithmic}
\end{algorithm}

\paragraph{Step 3: Cone separation and rank-two rescaling (\Cref{alg:private-perceptron-informal}).}
\Cref{alg:private-perceptron-informal} summarizes our solver, where $\widetilde a_j$ ranges over all private rows and public separators used during the epoch with weights $\widetilde\lambda_j$ summing to $1$ so that $\mu=\sum_j\widetilde\lambda_j\widetilde a_j$ is the noiseless average step; see \Cref{app:mixed-solver} for the full specification. Two modifications extend the framework to general cones and eliminate both bottlenecks (a) and (b):
\begin{enumerate}[label=\emph{\arabic*.}]
\item \emph{Public cone separation steps.}
Because a non-polyhedral cone $C_K$ has infinitely many supporting halfspaces, we enforce membership in $C_K$ via its separation oracle: whenever $y\notin C_K$, the oracle returns a unit vector $p\in C_K^*$ satisfying $\langle p,y\rangle\le 0$. Since $p$ depends only on $C_K$ and the public iterate $y$, this step incurs zero privacy loss by post-processing, while preserving $\|g\|_2\le 1$ and $\langle g,y\rangle\le 0$ for the perceptron drift analysis; adding fresh Gaussian noise $\zeta$ keeps the martingale structure uniform across all steps in \Cref{lem:support-function-bound}. Moreover, halting only when $y\in C_K$ guarantees $\widehat x\in K$.

\item \emph{Rank-two rescaling and directional noise averaging.}
Because both the violated private rows and the separators $p\in C_K^*$ are nonnegative on $F$, we build the noisy halfspace sub-sum $\widehat c$ from the combined family $\{\widetilde a_j\}$ and pair it with the noisy average step $\widehat\mu=y/T$ via the rank-two map
\begin{align*}
M
&:= \Bigl(I+\frac{\widehat c\,\widehat c^\top-4\widehat\mu\widehat\mu^\top}{\|\widehat c\|_2^2}\Bigr)^{1/2}.
\end{align*}
In the noiseless case ($\widehat c=c$ and $\widehat\mu=\mu$), every $z\in F$ satisfies $0\le\langle c,z\rangle\le\langle\mu,z\rangle$ and hence
\begin{align*}
\|Mz\|_2^2
&= \|z\|_2^2 + \frac{\langle c,z\rangle^2 - 4\langle\mu,z\rangle^2}{\|c\|_2^2}
\le \|z\|_2^2,
\end{align*}
so contraction along $\mu$ automatically compensates for expansion along $c$ across all of $F$, mapping $F\cap\mathbb B^d$ into $\mathbb B^d$ without needing $\|\mu\|_2=O(1/d)$. Furthermore, $\det M\ge\sqrt{3/2}$ holds as soon as $\|\widehat\mu\|_2\le\|\widehat c\|_2/4$; since accepted directions satisfy $\|\widehat c\|_2=\Omega(1/\sqrt{d})$, this requires only $\|\widehat\mu\|_2=O(1/\sqrt{d})$, cutting the epoch length to $T=\Theta(d)$ and removing bottleneck (a).

To handle the rescaling noise $\gamma\sim\mathcal N(0,\theta^2 I)$ in $\widehat c$ (and similarly in $\widehat\mu$), \Cref{lem:volume-increase} applies Jensen's inequality over $\omega$ drawn uniformly from $F\cap S^{d-1}$:
\begin{align*}
\mathrm{vol}(MF\cap\mathbb B^d)
&\ge \det M \cdot \bigl(\mathbb E_\omega[\|M\omega\|_2^2]\bigr)^{-d/2}\,\mathrm{vol}(F\cap\mathbb B^d).
\end{align*}
Consequently, $\gamma$ enters the volume factor only through the directional average $\mathbb E_\omega[\langle\gamma,\omega\rangle^2]=\Theta(\theta^2)$ rather than $\|\gamma\|_2^2=\Theta(d\theta^2)$. Tolerating $\sqrt{d}$ times larger noise lowers each rescaling release's privacy cost from $d\varepsilon_1$ to $\sqrt{d}\,\varepsilon_1$, removing bottleneck (b). With $O\bigl(d^2\log(1/\rho)\bigr)$ noisy steps and $\widetilde O\bigl(d\log(1/\rho)\bigr)$ rescaling trials, both compositions now permit $\varepsilon_1=\widetilde\Theta\bigl(\varepsilon/(d\sqrt{\log(1/\rho)})\bigr)$, yielding $\nu=\widetilde O\bigl((d^{3/2}/\varepsilon)\sqrt{\log(1/\rho)}\bigr)$ and proving \Cref{thm:upper-informal} upon substituting $d=k+1$ and our margin bound.
\end{enumerate}

\subsection{Lower bound}
\label{subsec:main-lower}

\begin{theorem}[Lower bound, informal version of \Cref{thm:lower-bound-main} and \Cref{cor:lower-add-remove,cor:lower-lorentz-psd}]
\label{thm:lower-informal}
Let $\varepsilon\in(0,1]$, $U>0$, $k\ge 2$, $R\in(0,k]$, $\eta\in(0,UR/(Ck)]$, and $\delta\in[0,\varepsilon/C]$, where $C$ is a suitable absolute constant.  Then there is a family of $R$-feasible LP instances over $\R_+^k$ (\Cref{def:problem}) such that, if an $(\varepsilon,\delta)$-DP algorithm under add/remove row adjacency satisfies $\Pr[\viol_\eta(\widehat x)\le s]\ge 1/2$ on every instance of the family, then
$$
s\;=\;\Omega\!\left(\frac{k}{\varepsilon}\,\log\min\!\left\{\frac{UR}{k\eta},\ \frac{\varepsilon}{\delta}\right\}\right).
$$
With the same bound up to constant factors, this also holds when $\R_+^k$ is replaced by a single Lorentz cone or a PSD cone of ambient dimension $k$ (non-polyhedral for $k\ge 3$).
\end{theorem}

Whenever $\eta\le UR/(Ck)$ and $\delta\le\varepsilon/C$, \Cref{thm:lower-informal} gives $s=\Omega(k/\varepsilon)$, showing that additive slack does not remove the linear dimension barrier proved for exact satisfaction by Kaplan et~al.~\cite{kaplan2024differentially}.  The condition $R\le k$ only keeps the right-hand sides of the hard instances in $[-U,U]$.  For $R>k$, the bound holds with $R$ replaced by $k$, since $k$-feasible instances are $R$-feasible.

\paragraph{Proof sketch.}
We pack one hidden symbol per coordinate into a narrow interval.  An output that $\eta$-satisfies most rows decodes most symbols, which group privacy rules out.

\medskip
\noindent
\emph{Hard instances.}
Set $\kappa:=\eta/U$, $M:=\lfloor UR/(64k\eta)\rfloor$, and symbol centers $\mu_m:=4\kappa m$ for $m\in[M]$.  For a hidden word $w$ drawn uniformly from $[M]^k$ and a duplication count $t\ge 1$, the instance $D_t(w)$ over $\R_+^k$, in the variable $z\in\R^k$, consists of $t$ copies of each of the paired rows
$$
Uz_i\le U\mu_{w_i},
\qquad
-Uz_i\le -U\mu_{w_i},
\qquad i\in[k].
$$
Each of the $2kt$ rows has norm $U$ and right-hand side in $[-U,U]$, and all hold with equality at $z^\star(w)=(\mu_{w_1},\ldots,\mu_{w_k})\in\R_+^k$, which has $\|z^\star(w)\|_2\le R$ by the choice of $M$.

\medskip
\noindent
\emph{Decoding.}
Both rows of coordinate $i$ are $\eta$-satisfied at $\widehat x$ if and only if $|\widehat x_i-\mu_{w_i}|\le\kappa$.  Since distinct centers are at least $4\kappa$ apart, the intervals $[\mu_m-\kappa,\mu_m+\kappa]$ are disjoint, so rounding $\widehat x_i$ to the nearest center recovers $w_i$.

\medskip
\noindent
\emph{Group privacy.}
Suppose for contradiction that $s<kt/2$.  Because all $t$ copies of a row are $\eta$-violated together, an output with at most $s$ violations leaves more than $k/2$ coordinates with both rows $\eta$-satisfied.  Such an output occurs with probability at least $1/2$, so for a uniformly random coordinate $I\in[k]$, rounding $\widehat x_I$ recovers $w_I$ with probability strictly above $1/4$.  Now replace $w_I$ by a fresh uniform symbol.  This changes only the $2t$ rows of coordinate $I$, at most $4t$ add/remove steps, and the original symbol no longer enters the data, so it is independent of the output and rounding recovers it with probability exactly $1/M$.  Group privacy (\Cref{fact:group-privacy}), with additive term at most $\delta e^{4t\varepsilon}/(e^\varepsilon-1)$, then gives
$$
\frac14 \;<\; e^{4t\varepsilon}\!\left(\frac{1}{M}+\frac{\delta}{e^\varepsilon-1}\right).
$$
Take $t=\lfloor c\varepsilon^{-1}\log\min\{M,\,\varepsilon/\delta\}\rfloor$ for a small constant $c>0$, so that $t\ge 1$ when $C$ is large.  This choice makes the right-hand side at most $1/4$, a contradiction.  Hence $s\ge kt/2$, which is $\Omega\bigl((k/\varepsilon)\log\min\{UR/(k\eta),\,\varepsilon/\delta\}\bigr)$.  Here $\log M=\log\Theta(UR/(k\eta))$ is the logarithm of the alphabet size per coordinate, and $\varepsilon^{-1}\log(\varepsilon/\delta)$ bounds the number $t$ of copies before the $\delta$-term of group privacy dominates.

\medskip
\noindent
\emph{Lorentz and PSD cones.}
To extend the construction to a single Lorentz or PSD cone of ambient dimension $k$ (\Cref{cor:lower-lorentz-psd}), we replace the coordinate axes of $\R_+^k$ by $k-1$ orthonormal directions $F_1,\ldots,F_{k-1}$ orthogonal to a deep interior point $x_0$ of norm $R/\sqrt 2$ (for the PSD cone of $n\times n$ matrices, $x_0=\frac{R}{\sqrt{2n}}I_n$ and traceless directions), so that every witness $x_0+\sum_{i=1}^{k-1}\mu_{w_i}F_i$ lies in $K$ and has norm at most $R$.

\begin{remark}[Tightness for unbounded algorithms]
\label{rmk:lower-tight}
For computationally unbounded algorithms, the factor $k/\varepsilon$ in \Cref{thm:lower-informal} is tight for every $\delta\in[0,\varepsilon/C]$, and the matching upper bound below holds for every symmetric cone $K\subseteq\R^k$.  Fix a public $\eta/(2U)$-net $N\subseteq K\cap B_R(0)$.  A volume argument gives $\log|N|=O(k\log(1+UR/\eta))$.  Running the exponential mechanism~\cite{dwork2014algorithmic} over $N$ with score $-\viol_\eta(x)$ (sensitivity $1$) is $(\varepsilon,0)$-DP.  Since the net point nearest $x^\star$ has no $\eta$-violations, the output satisfies $$\viol_\eta(\widehat x)=O((\log|N|)/\varepsilon)=O\bigl((k/\varepsilon)\log(1+UR/\eta)\bigr)$$ with constant probability.  When $\delta\le\varepsilon k\eta/(UR)$ (as in pure DP) and $UR/\eta\ge k^2$, this also matches the logarithmic factor of \Cref{thm:lower-informal}.  For instance, both bounds are $\Theta(k\log k)$ at $\varepsilon=1$, $U=R=1$, and $\eta=\delta=k^{-2}$.
\end{remark}

\begin{remark}[A barrier to stronger lower bounds]
\label{rmk:np-barrier}
Directly sampling the exponential mechanism of \Cref{rmk:lower-tight} takes time exponential in $k$.  For LPs, however, sampling a grid point $x$ with probability proportional to $2^{m-\viol_\eta(x)}$ reduces to uniformly sampling certificates $(x,\omega)$, where $\omega\in\{0,1\}^m$ fixes $\omega_i=0$ on $\eta$-violated rows and leaves $\eta$-satisfied rows free.  Such $\NP$ witnesses can be generated uniformly with an $\NP$ oracle~\cite{bellare2000uniform}.  Hence, if $\NP\subseteq\BPP$, row-private LP feasibility admits a polynomial-time $(1,0)$-DP algorithm with $O(k\log(2+kUR/\eta))$ violations (\Cref{prop:np-bpp}).  Any unconditional lower bound above this rate for polynomial-time $(1,\delta)$-DP algorithms, such as $\omega(k\log k)$ violations at $U=R=1$ and $\eta=k^{-2}$, would therefore prove $\NP\not\subseteq\BPP$ (\Cref{cor:np-barrier}).
\end{remark}

\section{Limitations and Open Questions}
\label{sec:limitations}

Our results show that row-private SCP feasibility admits logarithmic dependence on the scale-to-slack ratio while requiring at least linear dependence on the ambient dimension $k$, leaving several open directions.
\paragraph{Dimension dependence.}
Closing the gap between our $\widetilde O(k^{3/2}/\varepsilon)$ upper bound and the $\Omega(k/\varepsilon)$ lower bound is our main open question.  Even for LPs, the best exponent of $k$ achievable in polynomial time lies between $1$ and $3/2$.  By \Cref{rmk:lower-tight}, unbounded algorithms achieve a linear dependence, and by \Cref{rmk:np-barrier}, a lower bound of $k^{1+\Omega(1)}$ for polynomial-time algorithms at $\varepsilon=1$ and scale-to-slack ratio polynomial in $k$ would prove $\NP\not\subseteq\BPP$.  An improved algorithm is therefore the more likely way to close the gap. In our analysis, the bound is a product of two factors: a noisy step in dimension $k$ needs the threshold $\nu\approx\sqrt k/\varepsilon_1$ to keep its noise below the step length, and composing $O(k^2\log(1/\rho))$ steps ($O(k\log(1/\rho))$ epochs of $\Theta(k)$ steps) forces $\varepsilon_1\approx\varepsilon/k$ (\Cref{subsec:main-upper}).  The $\widetilde O(k\log(1/\rho))$ rescaling releases, each at privacy level $\sqrt k\,\varepsilon_1$, force the same value of $\varepsilon_1$, so shortening the epochs further would not improve the bound.

\paragraph{Dependence on $\eta$, $\delta$, and cone geometry.}
Two further questions concern the dependence on $\eta$ and $\delta$ and the geometry of the cone.  First, for fixed $\delta>0$, our lower bound stops growing once $\eta<\delta UR/(\varepsilon k)$ and stays at $\Omega\bigl((k/\varepsilon)\log(\varepsilon/\delta)\bigr)$, whereas our bound and that of the exponential mechanism both keep growing as $\eta\to 0$.  The optimal joint dependence on $UR/\eta$ and $\varepsilon/\delta$ is open even for unbounded algorithms.  Second, our Lorentz and PSD lower bounds embed a cube around an interior point and do not use the curvature of the cone.  For unbounded algorithms, curvature can change the bound by at most a logarithmic factor (\Cref{rmk:lower-tight}).  We do not know whether it forces polynomial-time algorithms to violate more rows than on polyhedral cones.

\paragraph{Zero slack ($\eta=0$).}
For LPs with integer data, Cramer's rule guarantees a $2^{-O(k\log(kU))}$ margin gap at vertices, allowing~\cite{kaplan2024differentially,ene2025solving} to perturb every row slightly and privately eliminate tight constraints to achieve $\eta=0$.  For general SCPs, the iterated-squaring instance of \Cref{app:hardness-exact} shows that this perturb-and-eliminate scheme can require a doubly exponentially small perturbation ($2^{-2^{\Omega(k)}}$), making the rescaling perceptron's epoch bound $O(k\log(1/\rho_0))$ exponential in $k$.  \Cref{rmk:upper-slack} shows that a known positive lower bound on the margin at $\eta=0$ suffices for our algorithm.  We do not know whether a different polynomial-time method can achieve $\eta=0$ without such a bound.

\paragraph{Scope and model assumptions.}
Our violation bound grows as $k^{3/2}$, which is of order $n^3$ for $n\times n$ PSD matrices, and the algorithm makes $O(k^2\log(UR\sqrt r/\eta))$ separation-oracle calls.  Privacy holds for every input, but the utility guarantee needs public bounds $U$ and $R$, and overestimating them increases the bound only inside a logarithm.  Our running-time analysis uses the real-arithmetic model of~\cite{kaplan2024differentially}, with exact oracle answers and continuous noise, and we do not analyze a finite-precision implementation.  We treat feasibility only: a public linear objective with a known target value $v$ can be added as copies of the public row $\langle c,x\rangle\le v$, but we do not search for the optimal value privately or handle private objectives.

\section*{Acknowledgments}
A preliminary version of this paper appeared at NeurIPS 2026.  This version adds, with the help of the AI tool GPT-6 Astra, the rank-two rescaling that lowers the dimension factor of our violation bound from $k^2$ to $k^{3/2}$.  The authors take full responsibility for all results.

\bibliographystyle{alphaurl}
\bibliography{main}


\appendix
\crefalias{section}{appendix}
\crefalias{subsection}{subappendix}

\section{Private Rescaling Perceptron for Mixed Homogeneous Systems}
\label{app:mixed-solver}
In this appendix, we present and analyze a private rescaling perceptron for mixed homogeneous systems with positive margin, as defined in \Cref{def:mixed-margin}. The input consists of private unit rows $a_1,\ldots,a_m\in\R^d$ with $d\ge 2$ and a public closed convex cone $\mathcal K\subseteq\R^d$ accessed via a separation oracle. We review the required differential privacy tools and transformed separation oracles in \Cref{subsec:privacy-tools,subsec:sep-oracle}, describe our generalization of the rescaling perceptron of~\cite{ene2025solving} in \Cref{subsec:mixed-algorithm}, and establish its privacy and utility guarantees in \Cref{sec:privacy,sec:utility-mixed}.

\subsection{Differential privacy tools}
\label{subsec:privacy-tools}

We first review the standard mechanisms and heterogeneous composition bound~\cite{dwork2014algorithmic} used in our privacy proof. For a function $f$ mapping row lists to $\R^q$, its \emph{$\ell_p$-sensitivity} under add/remove adjacency is
\begin{align*}
\Delta_p f
&:= \sup_{D,D'\text{ neighbors}} \bigl\|f(D)-f(D')\bigr\|_p.
\end{align*}
We apply the Laplace mechanism calibrated to $\ell_1$-sensitivity for noisy violation counts, and the Gaussian mechanism calibrated to $\ell_2$-sensitivity for vector releases.

\begin{fact}[Laplace mechanism \cite{dwork2006calibrating,dwork2014algorithmic}]
\label{fact:laplace-mechanism}
Let $f$ have $\ell_1$-sensitivity $\Delta_1 f$, and let $\Lap(b)$ denote the Laplace distribution with scale $b$ and density $\tfrac{1}{2b}\exp(-|x|/b)$. For every $\varepsilon>0$, the mechanism
\begin{align*}
\M_{\mathrm{Lap}}(D) := f(D)+(Y_1,\ldots,Y_q),
\qquad
Y_1,\ldots,Y_q \stackrel{\mathrm{iid}}{\sim} \Lap\!\bigl(\Delta_1 f/\varepsilon\bigr),
\end{align*}
is $(\varepsilon,0)$-differentially private.
\end{fact}

\begin{fact}[Gaussian mechanism \cite{dwork2006distributed,dwork2014algorithmic}]
\label{fact:gaussian-mechanism}
Let $f$ have $\ell_2$-sensitivity $\Delta_2 f$. For every $\varepsilon\in(0,1)$ and $\delta\in(0,1)$, the mechanism
\begin{align*}
\M_{\mathrm{Gauss}}(D) := f(D)+\zeta,
\qquad
\zeta \sim \mathcal N(0,\sigma^2 I_q),
\end{align*}
with $\sigma\ge\Delta_2 f\cdot\sqrt{2\log(1.25/\delta)}/\varepsilon$, is $(\varepsilon,\delta)$-differentially private.
\end{fact}

\begin{fact}[Advanced composition {\cite{dwork2014algorithmic}; heterogeneous form \cite[Theorem~3.5]{kairouz2017composition}}]
\label{fact:advanced-composition}
Let $\M_1,\ldots,\M_L$ be adaptive mechanisms such that each $\M_i$ may depend on the outputs of $\M_1,\ldots,\M_{i-1}$ and is $(\varepsilon_i,\delta_i)$-DP for every fixed realization of those prior outputs, where $\varepsilon_i>0$ and $\delta_i\in[0,1]$ are deterministic parameters fixed in advance. Then for any $\delta'\in(0,1)$, the adaptive composition $(\M_1,\ldots,\M_L)$ is $\bigl(\varepsilon_{\mathrm{tot}},\,\sum_{i=1}^L\delta_i+\delta'\bigr)$-DP, where
\begin{align*}
\varepsilon_{\mathrm{tot}}
&= \sqrt{2\log(1/\delta')\sum_{i=1}^L \varepsilon_i^2} + \sum_{i=1}^L \varepsilon_i(e^{\varepsilon_i}-1).
\end{align*}
In particular, when $(\varepsilon_i,\delta_i)=(\varepsilon_0,\delta_0)$ for all $i\in[L]$, the composition is $\bigl(\sqrt{2L\log(1/\delta')}\,\varepsilon_0+L\varepsilon_0(e^{\varepsilon_0}-1),\,L\delta_0+\delta'\bigr)$-DP.
\end{fact}

The heterogeneous bound above is a slight weakening of~\cite[Theorem~3.5]{kairouz2017composition}, and its homogeneous specialization recovers the standard advanced composition theorem of~\cite{dwork2014algorithmic}. Throughout this section, we refer to the sequence of all noisy counts, iterates, and noisy rescaling vectors released so far, together with the public Gaussian vectors $h^{(s)}$, as the \emph{transcript}.

\subsection{Separation oracles under linear transformation}
\label{subsec:sep-oracle}

After each rescaling, \cref{alg:private-perceptron} operates on the transformed cone $B^{-1}\mathcal K$ for an invertible matrix $B\in\R^{d\times d}$. Because $(B^{-1}\mathcal K)^*=B^\top\mathcal K^*$, a single call to a separation oracle for $\mathcal K$ suffices to separate over $B^{-1}\mathcal K$: a point $x$ belongs to $B^{-1}\mathcal K$ if and only if $Bx\in\mathcal K$, and whenever $Bx\notin\mathcal K$, any separator $q\in\mathcal K^*$ satisfying $\langle q,Bx\rangle<0$ maps to $B^\top q\in(B^{-1}{\cal K})^*=B^\top {\cal K}^*$, which satisfies
\begin{align*}
\langle B^\top q, x\rangle
&= \langle q, Bx\rangle < 0
\end{align*}
and is then normalized to unit length.

\subsection{Algorithm}
\label{subsec:mixed-algorithm}

\cref{alg:private-perceptron} forms the outer loop, running for at most $\tau=\lfloor d\log(1/\rho_0)/\log 1.1\rfloor+1$ epochs; each epoch invokes $\PPE$ in~\cref{alg:private-perceptron-epoch} and either returns a feasible point, applies a rescaling update, or aborts.

The epoch subroutine $\PPE$ departs from that of~\cite{ene2025solving} in three respects. 
\begin{itemize}
    \item First, during the perceptron phase, whenever $x^{(t)}\notin\mathcal K$, the algorithm queries the separation oracle for a unit separator $p^{(t)}\in\mathcal K^*$ satisfying $\langle p^{(t)},x^{(t)}\rangle\le 0$ and takes $g^{(t)}=p^{(t)}$ as the update direction. Fresh Gaussian noise $\zeta^{(t)}$ is still added on separator steps so that every iterate obeys the uniform recursion $x^{(t+1)}=x^{(t)}+g^{(t)}+\zeta^{(t)}$ used in the martingale analysis of \Cref{lem:support-function-bound}. 
    \item Second, once the perceptron phase finishes, the public separators collected during the epoch are appended to the normalized private rows with weight $1/T$ per occurrence, and the rescaling phase forms Gaussian halfspace sums over this augmented family.
    \item Third, the rescaling phase incorporates the noisy average step $\widehat\mu=x^{(T+1)}/T$, which is already determined by the transcript. A trial $s$ succeeds when its noisy subset sum satisfies $\|\widehat c^{(s)}\|_2\ge 1/(16\sqrt{\pi d})$ and $\|\widehat\mu\|_2\le\|\widehat c^{(s)}\|_2/4$, whereupon~\cref{alg:private-perceptron} applies the \emph{rank-two rescaling} $M$ that stretches along $\widehat c^{(s)}$ while contracting along $\widehat\mu$. Because $M$ acts as the identity on $\operatorname{span}\{\widehat c^{(s)},\widehat\mu\}^\perp$, computing $M$ and $M^{-1}$ reduces to a $2\times 2$ eigendecomposition. As we show in \Cref{sec:utility-mixed}, this rank-two update reduces the required epoch length from $T=\Theta(d^2)$ in~\cite{ene2025solving} to $T=\Theta(d)$, while allowing a rescaling noise scale $\theta$ that is $\sqrt{d}$ times larger.
\end{itemize}

\begin{algorithm}[tbp]
\caption{Private Rescaling Perceptron for a Mixed Homogeneous System}
\label{alg:private-perceptron}
\begin{algorithmic}[1]
\State \textbf{Input:} private unit rows $A\in\R^{m\times d}$, public cone $\mathcal K$, margin lower bound $\rho_0$, privacy $(\varepsilon,\delta)$, failure probability $\beta$
\State $\beta_1\gets\beta/4$, $\tau\gets\lfloor d\log(1/\rho_0)/\log 1.1\rfloor+1$, $T\gets\lceil(768\sqrt\pi)^2d\rceil$, $R_{\mathrm{try}}\gets\lceil 1000\log(\tau/\beta_1)\rceil$
\State $M_p\gets\tau T$, $M_r\gets\tau R_{\mathrm{try}}$, $M_0\gets M_p+M_r$, $Q\gets 4M_p+dM_r$
\State $\delta'\gets\delta/2$, $\delta_1\gets\delta/(2M_0)$, $\varepsilon_1\gets\varepsilon/\bigl(8\sqrt{Q\log(2/\delta)}\bigr)$
\State $\nu\gets\Theta\!\left(\frac{\sqrt d}{\varepsilon_1}\log^{3/2}\frac{T\tau}{\beta_1\delta_1}\right)$
\State $\sigma^2\gets 8\log(4T/\delta_1)/(\nu^2\varepsilon_1^2)$
\State $\theta^2\gets \frac{8}{d\,\varepsilon_1^2\nu^2}\cdot\log\bigl(4000\log(\tau/\beta_1)/\delta_1\bigr)$
\State $B\gets I$
\For{$\ell=1,\ldots,\tau$}
    \State $\mathrm{out}\gets \PPE(A,B^{-1}\mathcal K;\varepsilon_1,\delta_1,\beta_1,\nu,\sigma,\theta,T,R_{\mathrm{try}})$\Comment{\cref{alg:private-perceptron-epoch}}
    \If{$\mathrm{out}$ is a solution $z$}
        \State \Return $Bz$
    \ElsIf{$\mathrm{out}$ is a rescaling pair $(\widehat c,\widehat\mu)$}
        \State $M\gets\bigl(I+(\widehat c\,\widehat c^\top-4\widehat\mu\widehat\mu^\top)/\|\widehat c\|_2^2\bigr)^{1/2}$ \Comment{rank-two rescaling}
        \State $A \gets AM^{-1}$
        \State $B \gets BM^{-1}$
    \Else
        \State \Return $\bot$
    \EndIf
\EndFor
\State \Return $\bot$
\end{algorithmic}
\end{algorithm}

\begin{algorithm}[tbp]
\caption{$\PPE(A,\mathcal K;\varepsilon_1,\delta_1,\beta_1,\nu,\sigma,\theta,T,R_{\mathrm{try}})$}
\label{alg:private-perceptron-epoch}
\begin{algorithmic}[1]
\State Let $\bar a_1,\ldots,\bar a_m$ be the normalized rows of $A$
\State $x^{(1)}\gets 0\in\R^d$, $\lambda^{(1)}\gets0\in\R^m$
\For{$t=1,\ldots,T$} \Comment{perceptron phase}
    \If{$x^{(t)}\notin\mathcal K$}
        \State Let $p^{(t)}\in\mathcal K^*$ be a unit separator with $\langle p^{(t)},x^{(t)}\rangle\le0$
        \State $g^{(t)}\gets p^{(t)}$, $\lambda^{(t+1)}\gets\lambda^{(t)}$
    \Else
        \State $S^{(t)}\gets\{i\in[m]:\langle\bar a_i,x^{(t)}\rangle\le0\}$, $N^{(t)}\gets|S^{(t)}|$
        \State $\widehat N^{(t)}\gets N^{(t)}+\Lap(1/\varepsilon_1)-\varepsilon_1^{-1}\log\frac{2TR_{\mathrm{try}}}{\delta_1}$
        \If{$\widehat N^{(t)}\le\nu$}
            \State \Return solution $x^{(t)}$
        \EndIf
        \If{$N^{(t)}=0$} \Comment{avoids division by zero; see~\cref{prop:privacy-composition}}
            \State \Return solution $x^{(t)}$ \Comment{$x^{(t)}$ violates no row}
        \EndIf
        \State $u^{(t)}\gets \frac1{N^{(t)}}\sum_{i\in S^{(t)}}\bar a_i$, $g^{(t)}\gets u^{(t)}$
        \State $\lambda^{(t+1)}\gets\lambda^{(t)}$; for every $i\in S^{(t)}$, $\lambda_i^{(t+1)}\gets\lambda_i^{(t+1)}+1/N^{(t)}$
    \EndIf
    \State Sample $\zeta^{(t)}\sim \mathcal N(0,\sigma^2I)$; $x^{(t+1)}\gets x^{(t)}+g^{(t)}+\zeta^{(t)}$
\EndFor
\State $J_{\mathrm{pub}}\gets\{t\in[T]:x^{(t)}\notin\mathcal K\}$
\State Form the augmented row family $\widetilde A=(\widetilde a_j)_j:=(\bar a_1,\ldots,\bar a_m,(p^{(t)})_{t\in J_{\mathrm{pub}}})$
\State Form weights $\widetilde\lambda=(\lambda_1^{(T+1)}/T,\ldots,\lambda_m^{(T+1)}/T,(1/T)_{t\in J_{\mathrm{pub}}})$
\State $\widehat\mu\gets x^{(T+1)}/T$ \Comment{noisy average of the steps}
\For{$s=1,\ldots,R_{\mathrm{try}}$} \Comment{rescaling phase}
    \State Sample $h^{(s)}\sim \mathcal N(0,I)$
    \State $\widetilde P^{(s)}\gets\{j:\langle\widetilde a_j,h^{(s)}\rangle\ge0\}$
    \State $c^{(s)}\gets\sum_{j\in\widetilde P^{(s)}}\widetilde\lambda_j\widetilde a_j$
    \State Sample $\gamma^{(s)}\sim \mathcal N(0,\theta^2I)$; $\widehat c^{(s)}\gets c^{(s)}+\gamma^{(s)}$
    \If{$\|\widehat c^{(s)}\|_2\ge 1/(16\sqrt{\pi d})$ and $\|\widehat\mu\|_2\le\|\widehat c^{(s)}\|_2/4$}
        \State \Return rescaling pair $(\widehat c^{(s)},\widehat\mu)$
    \EndIf
\EndFor
\State \Return $\bot$
\end{algorithmic}
\end{algorithm}

\subsection{Privacy analysis}
\label{sec:privacy}

\begin{proposition}[Privacy of~\cref{alg:private-perceptron}]
\label{prop:privacy-composition}
Let $\varepsilon,\delta\in(0,1)$, and let $M_p=\tau T$, $M_r=\tau R_{\mathrm{try}}$, and $M_0=M_p+M_r$. Any choice of parameters $\varepsilon_1,\delta_1,\delta'>0$ satisfying
\begin{gather*}
\sqrt{2M_p\log\frac{1}{\delta'}}\,(2\varepsilon_1) + M_p\,(2\varepsilon_1)(e^{2\varepsilon_1}-1) \le \frac{\varepsilon}{2}, \\
\sqrt{2M_r\log\frac{1}{\delta'}}\,(\sqrt{d}\,\varepsilon_1) + M_r\,(\sqrt{d}\,\varepsilon_1)(e^{\sqrt{d}\,\varepsilon_1}-1) \le \frac{\varepsilon}{2}, \\
M_0\delta_1 + \delta' \le \delta
\end{gather*}
guarantees~\cref{alg:private-perceptron} $(\varepsilon,\delta)$-differentially private under add/remove adjacency. 

In particular, the parameter choices in~\cref{alg:private-perceptron} satisfy all three conditions and achieve $\varepsilon_1=\widetilde\Theta\bigl(\varepsilon/(d\sqrt{\log(1/\rho_0)})\bigr)$.
\end{proposition}

\begin{proof}
Let $\M$ denote~\cref{alg:private-perceptron}. Directly analyzing $\M$ faces a technical obstacle: on a perceptron step inside $\mathcal K$, the update $u^{(t)}=\frac{1}{N^{(t)}}\sum_{i\in S^{(t)}}\bar a_i$ divides by the true violation count $N^{(t)}$ and includes an unnoised guard for $N^{(t)}=0$. Although the stopping test $\widehat N^{(t)}\le\nu$ is designed to halt whenever $N^{(t)}<\nu$, a large positive Laplace draw could cause $\M$ to continue when $N^{(t)}$ is small, ruining the worst-case $\ell_2$-sensitivity of $u^{(t)}$. To handle this cleanly, our proof proceeds in four steps:
\begin{enumerate}
\item We define a clipped surrogate $\M'$ that omits the $N^{(t)}=0$ check and replaces $N^{(t)}$ by $\max\{N^{(t)},\nu\}$ in all denominators, guaranteeing $O(1/\nu)$ sensitivity unconditionally, while checking that all public-cone operations depend only on the released transcript.
\item We prove per-slot privacy for $\M'$: separator steps cost $(0,0)$-DP by post-processing, private perceptron steps satisfy $(2\varepsilon_1,\delta_1/2)$-DP, and rescaling trials satisfy $(\sqrt{d}\,\varepsilon_1,\delta_1/2)$-DP. Spending a $\sqrt{d}$ larger privacy budget on rescaling trials shrinks the noise scale $\theta$ by $\sqrt{d}$ at no asymptotic cost because rescaling trials are $\widetilde\Theta(d)$ times rarer than perceptron steps.
\item We pad $\M'$ to $M_0=M_p+M_r$ deterministic slots and apply heterogeneous advanced composition (\Cref{fact:advanced-composition}) to conclude that $\M'$ is $(\varepsilon,\tfrac{1}{2}M_0\delta_1+\delta')$-DP.
\item We couple $\M$ and $\M'$ under shared randomness: because $\widehat N^{(t)}$ shifts $N^{(t)}$ downward, the two algorithms agree unless a Laplace variable exceeds that shift, which happens with probability much smaller than $\tfrac{1}{2}M_0\delta_1$ and is absorbed into $\delta$.
\end{enumerate}

\emph{Step 1: Transcript invariants and the clipped surrogate $\M'$.}
Call a perceptron step \emph{private} if $x^{(t)}\in\mathcal K$ and a \emph{separator step} otherwise. Conditioned on the transcript at the start of an epoch, the active rows are obtained from the input rows via a fixed invertible linear map followed by row-wise normalization, so add/remove neighboring inputs induce neighboring active row lists. Within an epoch, whether $x^{(t)}\in\mathcal K$, the chosen separator $p^{(t)}$, and the index set $J_{\mathrm{pub}}$ depend solely on the public cone $\mathcal K$ and the public iterates $x^{(t)}$. Likewise, $\widehat\mu=x^{(T+1)}/T$ is a function of the transcript, and~\cref{alg:private-perceptron-epoch} returns a rescaling pair $(\widehat c^{(s)},\widehat\mu)$ only when $\|\widehat c^{(s)}\|_2\ge 1/(16\sqrt{\pi d})>0$ and $\|\widehat\mu\|_2\le\|\widehat c^{(s)}\|_2/4$; by \Cref{lem:volume-increase}, every accepted rescaling matrix $M$ is therefore well defined and invertible.

Let $\M'$ be identical to $\M$ except that it omits the $N^{(t)}=0$ stopping rule and replaces $1/N^{(t)}$ with $1/\max\{N^{(t)},\nu\}$ in both $u^{(t)}$ and the increments of $\lambda$, giving $u^{(t)}=0$ if $N^{(t)}=0$. Under $\M'$, whether a private step continues is determined purely by the public test $\widehat N^{(t)}>\nu$.

\emph{Step 2: Conditional sensitivity and per-slot privacy of $\M'$.}
Note first that the second condition in~\cref{prop:privacy-composition} implies $\sqrt{d}\,\varepsilon_1(e^{\sqrt{d}\,\varepsilon_1}-1)<1/2$, whence $\varepsilon_1\le\sqrt{d}\,\varepsilon_1<1$, so \Cref{fact:gaussian-mechanism} applies at both privacy levels $\varepsilon_1$ and $\sqrt{d}\,\varepsilon_1$.

Both the clipped perceptron direction $u^{(t)}$ and each private summand of a rescaling vector $c^{(s)}$ take the form $\frac{1}{\max\{N^{(t)},\nu\}}\sum_{i\in V}\bar a_i$ for a subset $V\subseteq S^{(t)}$, so we bound their sensitivities simultaneously. Fix the transcript up to step $t$ and consider neighboring active row lists $D$ and $D'$, where $D'$ adds one unit row $\bar a$. If $\bar a$ is not violated at $x^{(t)}$, then $S^{(t)}$ and all step-$t$ quantities are identical on $D$ and $D'$. Otherwise, let $N:=|S^{(t)}(D)|$, so $|S^{(t)}(D')|=N+1$, and let $n:=\max\{N,\nu\}$ and $n':=\max\{N+1,\nu\}$, which satisfy $n\le n'\le n+1$. For any subset $V\subseteq S^{(t)}(D)$ of common violated rows and any indicator $\chi\in\{0,1\}$, we have $|V|\le N\le n$, and thus
\begin{align*}
\left\|\Bigl(\frac{1}{n'}-\frac{1}{n}\Bigr)\sum_{i\in V}\bar a_i + \frac{\chi\,\bar a}{n'}\right\|_2
&\le \frac{|V|\,(n'-n)}{n\,n'} + \frac{1}{n'}
\le \frac{2}{n'}
\le \frac{2}{\nu},
\end{align*}
where the first inequality applies the triangle inequality to the unit vectors $\bar a_i$ and $\bar a$, the second uses $|V|\le n$ and $n'-n\le 1$, and the third uses $n'\ge\nu$. We now deduce the privacy cost of each step type in $\M'$:
\begin{itemize}
\item \emph{Private perceptron steps.} Taking $V=S^{(t)}(D)$ and $\chi=1$ bounds the conditional $\ell_2$-sensitivity of $u^{(t)}$ by $2/\nu$. Since $N^{(t)}$ has $\ell_1$-sensitivity $1$, releasing $\widehat N^{(t)}$ is $(\varepsilon_1,0)$-DP by \Cref{fact:laplace-mechanism}, and if $\widehat N^{(t)}>\nu$, releasing $x^{(t+1)}=x^{(t)}+u^{(t)}+\zeta^{(t)}$ is $(\varepsilon_1,\delta_1/2)$-DP by \Cref{fact:gaussian-mechanism} and our choice of $\sigma$. Basic composition shows that each private step is $(2\varepsilon_1,\delta_1/2)$-DP, matching Lemma~6 of Ene et~al.~\cite{ene2025solving}.

\item \emph{Separator steps.} When $x^{(t)}\notin\mathcal K$, the step releases $x^{(t+1)}=x^{(t)}+p^{(t)}+\zeta^{(t)}$ without touching the private rows, since $p^{(t)}$ depends only on $\mathcal K$ and $x^{(t)}$. Conditioned on the transcript, this step is $(0,0)$-DP by post-processing.

\item \emph{Rescaling trials.} In trial $s$, the Gaussian vector $h^{(s)}$ is public and $c^{(s)}=c^{(s)}_{\mathrm{pub}}+c^{(s)}_{\mathrm{priv}}$, where
\begin{align*}
c^{(s)}_{\mathrm{pub}}
:= \frac{1}{T}\sum_{\substack{t\in J_{\mathrm{pub}}:\\ \langle p^{(t)},h^{(s)}\rangle\ge 0}} p^{(t)},\qquad
c^{(s)}_{\mathrm{priv}}:= \sum_{t\in[T]\setminus J_{\mathrm{pub}}} \frac{1}{T\max\{N^{(t)},\nu\}} \sum_{i\in S^{(t)}\cap\widetilde P^{(s)}} \bar a_i,
\end{align*}
with the second sum running over continued private steps. Conditioned on the transcript and $h^{(s)}$, the public part $c^{(s)}_{\mathrm{pub}}$ is fixed, while applying our sensitivity estimate to each of the at most $T$ summands of $c^{(s)}_{\mathrm{priv}}$ with $V=S^{(t)}(D)\cap\widetilde P^{(s)}$ and $\chi=\mathbf{1}\{\langle\bar a,h^{(s)}\rangle\ge 0\}$ bounds each summand's change by $2/(T\nu)$. Hence $\Delta_2 c^{(s)}\le 2/\nu$, as in~\cite[Claim~9]{ene2025solving}. By \Cref{fact:gaussian-mechanism} at privacy parameter $\sqrt{d}\,\varepsilon_1$ and our choice of $\theta$, releasing $\widehat c^{(s)}$ is $(\sqrt{d}\,\varepsilon_1,\delta_1/2)$-DP, matching~\cite[Lemma~8]{ene2025solving}.
\end{itemize}

\emph{Step 3: Heterogeneous composition for $\M'$.}
Pad every run of $\M'$ to $M_0=M_p+M_r$ fixed slots: $T$ perceptron slots followed by $R_{\mathrm{try}}$ rescaling slots in each of the $\tau$ epochs, and emitting a fixed dummy symbol in any slot after an epoch or the algorithm has halted. Because $(0,0)$-DP implies $(2\varepsilon_1,\delta_1/2)$-DP, every perceptron slot is $(2\varepsilon_1,\delta_1/2)$-DP conditioned on the prior transcript regardless of whether it executes a private step, a separator step, or a dummy output; similarly, every rescaling slot is $(\sqrt{d}\,\varepsilon_1,\delta_1/2)$-DP. Since these parameters depend only on the slot index, \Cref{fact:advanced-composition} implies that the padded transcript is $(\varepsilon_{\mathrm{tot}},\,\tfrac{1}{2}M_0\delta_1+\delta')$-DP, where
\begin{align*}
\varepsilon_{\mathrm{tot}}
&= \sqrt{2\log\frac{1}{\delta'}\Bigl(M_p(2\varepsilon_1)^2 + M_r(\sqrt{d}\,\varepsilon_1)^2\Bigr)}
   + M_p(2\varepsilon_1)(e^{2\varepsilon_1}-1)
   + M_r(\sqrt{d}\,\varepsilon_1)(e^{\sqrt{d}\,\varepsilon_1}-1) \\
&\le \left(\sqrt{2M_p\log\tfrac{1}{\delta'}}\,(2\varepsilon_1) + M_p(2\varepsilon_1)(e^{2\varepsilon_1}-1)\right)
   + \left(\sqrt{2M_r\log\tfrac{1}{\delta'}}\,(\sqrt{d}\,\varepsilon_1) + M_r(\sqrt{d}\,\varepsilon_1)(e^{\sqrt{d}\,\varepsilon_1}-1)\right) \\
&\le \frac{\varepsilon}{2} + \frac{\varepsilon}{2}
= \varepsilon.
\end{align*}
Here, the first inequality applies $\sqrt{a+b}\le\sqrt{a}+\sqrt{b}$ to decouple the perceptron and rescaling contributions, and the second applies the first two conditions of the proposition. Because the output of $\M'$ is a post-processing of the padded transcript, $\M'$ is $(\varepsilon,\,\tfrac{1}{2}M_0\delta_1+\delta')$-DP.

\emph{Step 4: Coupling $\M$ to $\M'$.}
Finally, couple $\M$ and $\M'$ on any input $D$ using shared internal randomness. The two executions produce identical transcripts and outputs unless some private step satisfies $\widehat N^{(t)}>\nu$ while $N^{(t)}<\nu$: prior to the first such step, passing $\widehat N^{(t)}>\nu$ implies $N^{(t)}\ge\nu>0$, so the $N^{(t)}=0$ guard in $\M$ never triggers and $\max\{N^{(t)},\nu\}=N^{(t)}$ holds on every continued step. Conditioned on the history at any step where $N^{(t)}<\nu$, the event $\widehat N^{(t)}>\nu$ requires the Laplace draw $Y\sim\Lap(1/\varepsilon_1)$ to satisfy
\begin{align*}
Y
&= \widehat N^{(t)} - N^{(t)} + \frac{1}{\varepsilon_1}\log\frac{2TR_{\mathrm{try}}}{\delta_1}
> \frac{1}{\varepsilon_1}\log\frac{2TR_{\mathrm{try}}}{\delta_1},
\end{align*}
which occurs with probability $\tfrac{1}{2}\exp\bigl(-\log(2TR_{\mathrm{try}}/\delta_1)\bigr)=\delta_1/(4TR_{\mathrm{try}})$. A union bound over all at most $\tau T$ private steps gives $\Pr[\M(D)\ne\M'(D)]\le\delta_2:=\tau\delta_1/(4R_{\mathrm{try}})$ on every input $D$. Therefore, for any neighboring inputs $D,D'$ and any measurable set $S$,
\begin{align*}
\Pr[\M(D)\in S]
&\le \Pr[\M'(D)\in S] + \delta_2 \\
&\le e^{\varepsilon}\Pr[\M'(D')\in S] + \frac{1}{2}M_0\delta_1 + \delta' + \delta_2 \\
&\le e^{\varepsilon}\Pr[\M(D')\in S] + \frac{1}{2}M_0\delta_1 + \delta' + (1+e^{\varepsilon})\delta_2,
\end{align*}
where the first and third inequalities apply the coupling bound on $D$ and $D'$, and the second applies $(\varepsilon,\tfrac{1}{2}M_0\delta_1+\delta')$-DP of $\M'$. Since $\varepsilon<1$, $R_{\mathrm{try}}\ge 1$, and $T+R_{\mathrm{try}}\ge 2$, we have $(1+e^{\varepsilon})\delta_2<\tau\delta_1\le\tfrac{1}{2}M_0\delta_1$. Combining this with the third condition $M_0\delta_1+\delta'\le\delta$ establishes that $\M$ is $(\varepsilon,\delta)$-DP.

\emph{Verification of the parameter choices in~\cref{alg:private-perceptron}.}
Let $Q:=4M_p+dM_r$ and $a:=\log(2/\delta)=\log(1/\delta')$, and consider the choices $\delta'=\delta/2$, $\delta_1=\delta/(2M_0)$, and $\varepsilon_1=\varepsilon/(8\sqrt{Qa})$ from Line~4 of~\cref{alg:private-perceptron}. We have $M_0\delta_1+\delta'=\delta$. Next, let $(N,q)$ be either $(M_p,2\varepsilon_1)$ or $(M_r,\sqrt{d}\,\varepsilon_1)$. In both cases $N\ge 1$ and $Nq^2\le(4M_p+dM_r)\varepsilon_1^2=Q\varepsilon_1^2=\varepsilon^2/(64a)$. Since $\varepsilon<1$ and $a>\log 2>1/2$, we have $q^2\le Nq^2<1/32$, so $q<1$ and $e^q-1\le 2q$. Consequently,
\begin{align*}
\sqrt{2N\log\frac{1}{\delta'}}\,q + Nq(e^q-1)
\le \sqrt{2Na\,q^2} + 2Nq^2 
\le \sqrt{2Qa\,\varepsilon_1^2} + 2Q\varepsilon_1^2 
= \frac{\sqrt{2}}{8}\,\varepsilon + \frac{\varepsilon^2}{32a}
< \frac{\varepsilon}{2},
\end{align*}
where the last step uses $\varepsilon<1$, $a>1/2$, and $\sqrt{2}/8+1/16<1/2$. Finally, since $M_p=\Theta\bigl(d^2\log(1/\rho_0)\bigr)$ and $M_r=\widetilde\Theta\bigl(d\log(1/\rho_0)\bigr)$, we have $Q=\widetilde\Theta\bigl(d^2\log(1/\rho_0)\bigr)$ and hence $\varepsilon_1=\widetilde\Theta\bigl(\varepsilon/(d\sqrt{\log(1/\rho_0)})\bigr)$.

\end{proof}

\subsection{Utility analysis}
\label{sec:utility-mixed}

Let $\mathbb B^d$ and $S^{d-1}$ denote the closed Euclidean unit ball and unit sphere in $\R^d$. Within a fixed epoch with current matrix $B$ and row matrix $A$, let $F:=\{x\in B^{-1}\mathcal K:Ax\ge 0\}$ be the mixed feasible cone and let $P:=F\cap\mathbb B^d$; inside \Cref{lem:support-function-bound}, we write $\mathcal K$ in place of $B^{-1}\mathcal K$ to match the notation of~\cref{alg:private-perceptron-epoch}. Following the volume-growth framework of~\cite{ene2025solving}, a mixed margin of $2\rho_0$ guarantees $\mathrm{vol}(P)\ge\rho_0^d\,\mathrm{vol}(\mathbb B^d)$ initially, while each rescaling increases $\mathrm{vol}(P)$ by a factor of at least $1.1$. Since $\mathrm{vol}(P)\le\mathrm{vol}(\mathbb B^d)$ throughout, at most $\lfloor d\log(1/\rho_0)/\log 1.1\rfloor=\tau-1$ rescalings can occur before an epoch terminates its perceptron phase with few violations.

We begin with three lemmas analyzing the rescaling phase, where \Cref{lem:volume-increase,lem:gaussian-subset} serve as rank-two and mixed-cone counterparts to~\cite[Lemmas~4 and~6]{hoberg2017improved}.

\begin{lemma}[Volume increase under the rank-two rescaling]
\label{lem:volume-increase}
Let $F\subseteq\R^d$ be a closed convex cone with nonempty interior, let $\omega$ be drawn uniformly from $F\cap S^{d-1}$, and let $\Sigma_F:=\mathbb E[\omega\omega^\top]$. Suppose $c,\mu\in\R^d$ satisfy $0\le\langle c,x\rangle\le\langle\mu,x\rangle$ for all $x\in F$, and let $\widehat c=c+\gamma$ and $\widehat\mu=\mu+\bar\xi$ for vectors $\gamma,\bar\xi\in\R^d$ satisfying $\widehat c\ne 0$ and $\|\widehat\mu\|_2\le\|\widehat c\|_2/4$. Then
\begin{align*}
M := \Bigl(I+\frac{\widehat c\,\widehat c^\top-4\widehat\mu\widehat\mu^\top}{\|\widehat c\|_2^2}\Bigr)^{1/2}
\end{align*}
is well defined and positive definite, and
\begin{align*}
\mathrm{vol}(MF\cap\mathbb B^d)
&\ge \sqrt{\frac{3}{2}}\,\exp\!\Bigl(-\frac{d\,(\gamma^\top\Sigma_F\gamma+2\bar\xi^\top\Sigma_F\bar\xi)}{\|\widehat c\|_2^2}\Bigr)\,\mathrm{vol}(F\cap\mathbb B^d).
\end{align*}
\end{lemma}

\begin{proof}
Let $u:=\widehat c/\|\widehat c\|_2$ and $w:=2\widehat\mu/\|\widehat c\|_2$, so that $\|u\|_2=1$, $\|w\|_2\le 1/2$, and $M^2=I+uu^\top-ww^\top$. For every unit vector $v\in S^{d-1}$, we have $v^\top M^2 v\ge 1-\|w\|_2^2\ge 3/4$, showing that $M$ is well defined and positive definite. Applying the matrix determinant lemma for rank-two updates gives
\begin{align*}
\det (M^2)
&= (1+\|u\|_2^2)(1-\|w\|_2^2) + \langle u,w\rangle^2
\ge 2\cdot\frac{3}{4}
= \frac{3}{2}.
\end{align*}
Changing variables via $y=Mx$, integrating in polar coordinates over the cone $F$, and applying Jensen's inequality to the convex map $s\mapsto s^{-d/2}$ on $(0,\infty)$ gives
\begin{align*}
\frac{\mathrm{vol}(MF\cap\mathbb B^d)}{\mathrm{vol}(F\cap\mathbb B^d)}
&= \det (M) \cdot \frac{\mathrm{vol}(\{x\in F:\|Mx\|_2\le 1\})}{\mathrm{vol}(F\cap\mathbb B^d)} \\
&= \det (M) \cdot \mathbb E\bigl[(\omega^\top M^2\omega)^{-d/2}\bigr] \\
&\ge \det (M) \cdot \bigl(\mathbb E[\omega^\top M^2\omega]\bigr)^{-d/2}.
\end{align*}
To upper-bound $\mathbb E[\omega^\top M^2\omega]$, fix any $x\in F$ and let $a:=\langle c,x\rangle$, $b:=\langle\mu,x\rangle$, $g:=\langle\gamma,x\rangle$, and $f:=\langle\bar\xi,x\rangle$, so that $0\le a\le b$ by hypothesis. Using the elementary inequalities $(a+g)^2\le 2a^2+2g^2$ and $(b+f)^2\ge b^2/2-f^2$, we obtain
\begin{align*}
\langle\widehat c,x\rangle^2 - 4\langle\widehat\mu,x\rangle^2
&= (a+g)^2 - 4(b+f)^2
\le 2(a^2-b^2) + 2g^2 + 4f^2
\le 2\langle\gamma,x\rangle^2 + 4\langle\bar\xi,x\rangle^2,
\end{align*}
where the last step uses $a^2\le b^2$. Evaluating at $x=\omega$, dividing by $\|\widehat c\|_2^2$, and taking expectations yields $\mathbb E[\omega^\top M^2\omega]\le 1+2z$, where $z:=(\gamma^\top\Sigma_F\gamma+2\bar\xi^\top\Sigma_F\bar\xi)/\|\widehat c\|_2^2$. Combining $(1+2z)^{-d/2}\ge e^{-dz}$ with $\det M\ge\sqrt{3/2}$ completes the proof.
\end{proof}

In the noiseless case where $\gamma=\bar\xi=0$, the pointwise inequality above gives $\|Mx\|_2\le\|x\|_2$ for all $x\in F$, so $M$ maps $F\cap\mathbb B^d$ into $\mathbb B^d$ while expanding volume by $\det (M)\ge\sqrt{3/2}$. Under noise, the perturbation enters only through the quadratic forms $\gamma^\top\Sigma_F\gamma$ and $\bar\xi^\top\Sigma_F\bar\xi$. Because $\operatorname{tr}\Sigma_F=\mathbb E\|\omega\|_2^2=1$, \Cref{lem:gaussian-quadratic} shows that $\gamma^\top\Sigma_F\gamma$ for $\gamma\sim\mathcal N(0,\theta^2 I)$ is of order $\theta^2$ rather than $\|\gamma\|_2^2=\Theta(d\theta^2)$, which explains why $\theta$ can be $\sqrt{d}$ times larger than in~\cite{ene2025solving}. Finally, updating $A\gets AM^{-1}$ and $B\gets BM^{-1}$ in~\cref{alg:private-perceptron} transforms the feasible cone $F$ into $MF$.

\begin{lemma}[Gaussian quadratic forms]
\label{lem:gaussian-quadratic}
Let $\Sigma\in\R^{d\times d}$ be positive semidefinite with $\operatorname{tr}\Sigma=1$, let $Z\sim\mathcal N(0,s^2I)$ with $s>0$, and let $p\in(0,1)$. Then $\Pr\bigl[Z^\top\Sigma Z\ge s^2\bigl(2+4\log(1/p)\bigr)\bigr]\le p$. 

In particular, taking $\Sigma=I/d$ yields $\Pr\bigl[\|Z\|_2^2\ge ds^2\bigl(2+4\log(1/p)\bigr)\bigr]\le p$.
\end{lemma}

\begin{proof}
Diagonalizing $\Sigma$ gives $Z^\top\Sigma Z=s^2\sum_{i=1}^d \sigma_i X_i^2$, where $\sigma_1,\ldots,\sigma_d\ge 0$ are the eigenvalues of $\Sigma$ summing to $1$ and $X_1,\ldots,X_d$ are independent standard Gaussians. Using $\mathbb E[e^{tX_i^2}]=(1-2t)^{-1/2}$ for $t<1/2$ together with $(1-y)^{-1/2}\le e^y$ for $y\in[0,1/2]$, we obtain
\begin{align*}
\mathbb E\exp\!\Bigl(\frac{Z^\top\Sigma Z}{4s^2}\Bigr)
&= \prod_{i=1}^d \Bigl(1-\frac{\sigma_i}{2}\Bigr)^{-1/2}
\le \exp\!\Bigl(\frac{1}{2}\sum_{i=1}^d \sigma_i\Bigr)
= e^{1/2}.
\end{align*}
Applying Markov's inequality to $\exp\bigl(Z^\top\Sigma Z/(4s^2)\bigr)$ then gives
\begin{align*}
\Pr\Bigl[Z^\top\Sigma Z \ge s^2\bigl(2+4\log(1/p)\bigr)\Bigr]
&\le e^{1/2}\,\exp\!\Bigl(-\frac{1}{2}-\log\frac{1}{p}\Bigr)
= p,
\end{align*}
which proves the lemma.
\end{proof}

\begin{lemma}[Gaussian subset sums; cf.\ {\cite[Lemma~6]{hoberg2017improved}}]
\label{lem:gaussian-subset}
Let $d\ge 2$, let $\widetilde a_1,\ldots,\widetilde a_n\in\R^d$ be unit vectors with nonnegative weights $\widetilde\lambda_j$ summing to $1$, let $h\sim\mathcal N(0,I)$, and let $c:=\sum_{j:\langle\widetilde a_j,h\rangle\ge 0}\widetilde\lambda_j\widetilde a_j$. Then $\Pr\bigl[\|c\|_2\ge 1/(8\sqrt{\pi d})\bigr]\ge 1/20$.
\end{lemma}

\begin{proof}
Let $X:=\langle c,h\rangle=\sum_{j=1}^n\widetilde\lambda_j\max\{\langle\widetilde a_j,h\rangle,0\}$. Since each $\langle\widetilde a_j,h\rangle\sim\mathcal N(0,1)$ and the weights $\widetilde\lambda_j$ form a convex combination, Jensen's inequality gives
\begin{align*}
\mathbb E[X]
&= \frac{1}{\sqrt{2\pi}},
&
\mathbb E[X^2]
&\le \sum_{j=1}^n \widetilde\lambda_j\,\mathbb E\bigl[\max\{\langle\widetilde a_j,h\rangle,0\}^2\bigr]
= \frac{1}{2}.
\end{align*}
Applying the Paley--Zygmund inequality to $X\ge 0$ and Markov's inequality to $\exp(3\|h\|_2^2/8)$, which has expectation $(1-3/4)^{-d/2}=2^d$, yields
\begin{align*}
\Pr\!\left[X \ge \frac{1}{2\sqrt{2\pi}}\right]
&\ge \frac{(\mathbb E X)^2}{4\,\mathbb E[X^2]}
\ge \frac{1}{4\pi},
&
\Pr\bigl[\|h\|_2^2 \ge 8d\bigr]
&\le (2e^{-3})^d
\le e^{-4},
\end{align*}
where the last step uses $d\ge 2$ and $2e^{-3}<e^{-2}$. Whenever $X\ge 1/(2\sqrt{2\pi})$ and $\|h\|_2<\sqrt{8d}$ both hold, Cauchy--Schwarz implies
\begin{align*}
\|c\|_2
&\ge \frac{X}{\|h\|_2}
\ge \frac{1}{2\sqrt{2\pi}\cdot\sqrt{8d}}
= \frac{1}{8\sqrt{\pi d}},
\end{align*}
which occurs with probability at least $1/(4\pi)-e^{-4}>1/20$.
\end{proof}

Next, we establish three auxiliary lemmas connecting the mixed margin and perceptron phase to the rescaling guarantees: \Cref{lem:initial-volume} lower-bounds the initial volume of $P$, \Cref{lem:support-function-bound} bounds the noisy average step $\widehat\mu$, and \Cref{fact:utility-stopping} bounds the violation count of any returned solution by a public threshold $\Gamma$, which \Cref{app:upper-proof} also uses to choose the multiplicity of $e_t$.

\begin{lemma}[Initial volume from the mixed margin]
\label{lem:initial-volume}
If $\rho(A,\mathcal K)\ge 2\rho_0$ with $\rho_0\le 1/2$, then $P$ contains a Euclidean ball of radius $\rho_0$, and consequently $\mathrm{vol}(P)\ge\rho_0^d\,\mathrm{vol}(\mathbb B^d)$.
\end{lemma}

\begin{proof}
Let $y\in\mathcal K$ with $\|y\|_2\le 1$ witness $\rho(A,\mathcal K)\ge 2\rho_0$, so that $\langle a_i,y\rangle\ge 2\rho_0$ for all $i\in[m]$ and $\langle h,y\rangle\ge 2\rho_0$ for every unit vector $h\in\mathcal K^*$. For any perturbation $u\in\R^d$ satisfying $\|u\|_2\le\rho_0$, Cauchy--Schwarz gives
\begin{align*}
\langle a_i,\, y/2+u\rangle
&\ge \rho_0 - \|u\|_2 \ge 0 \qquad \forall i\in[m], \\
\langle h,\, y/2+u\rangle
&\ge \rho_0 - \|u\|_2 \ge 0 \qquad \forall h\in\mathcal K^*\cap S^{d-1}.
\end{align*}
By the bipolar theorem, the second condition implies $y/2+u\in(\mathcal K^*)^*=\mathcal K$. Moreover, $\|y/2+u\|_2\le 1/2+\rho_0\le 1$, so the ball $y/2+\rho_0\mathbb B^d$ is contained in $P$, giving $\mathrm{vol}(P)\ge\rho_0^d\,\mathrm{vol}(\mathbb B^d)$.
\end{proof}

\begin{lemma}[Augmented family and average step]
\label{lem:support-function-bound}
In every epoch that reaches its rescaling phase, each augmented row $\widetilde a_j$ is a unit vector that is nonnegative on $F$, the weights $\widetilde\lambda_j$ are nonnegative and sum to $1$, and $\widehat\mu=\mu+\xi/T$ where $\mu:=\sum_j\widetilde\lambda_j\widetilde a_j$ and $\xi:=\sum_{t=1}^T\zeta^{(t)}$. Moreover, let $p\in(0,1)$ satisfy $\bigl(2+4\log(1/p)\bigr)d\sigma^2\le 1$. Conditioned on the history before an epoch, with probability at least $1-2p$, the epoch either terminates during its perceptron phase or reaches its rescaling phase with $\|\widehat\mu\|_2\le\sqrt{5/T}$.
\end{lemma}

\begin{proof}
By construction, every augmented row $\widetilde a_j$ has unit norm, and each of the $T$ perceptron steps contributes total weight $1/T$ to $\widetilde\lambda$, either split evenly across the $N^{(t)}$ violated rows when $x^{(t)}\in\mathcal K$ or assigned to the separator $p^{(t)}$ when $x^{(t)}\notin\mathcal K$; thus $\widetilde\lambda_j\ge 0$ and $\sum_j\widetilde\lambda_j=1$. Every private row is nonnegative on $F$ because $Ax\ge 0$ on $F$, and every separator $p^{(t)}\in\mathcal K^*$ is nonnegative on $F$ because $F\subseteq\mathcal K$. Furthermore, summing the deterministic step directions gives $\sum_{t=1}^T g^{(t)}=T\mu$, so $x^{(T+1)}=T\mu+\xi$ and $\widehat\mu=\mu+\xi/T$.

To bound $\|\widehat\mu\|_2$, we give a direct martingale argument extending~\cite[Claims~15 and~17]{ene2025solving} to accommodate separator steps. If the perceptron phase halts early, extend the sequence to $T$ steps by taking $g^{(t)}=0$ while still adding $\zeta^{(t)}$; this leaves unchanged both whether the epoch reaches the rescaling phase and the terminal iterate $x^{(T+1)}$ when it does. In every step $t\in[T]$, we then have $\|g^{(t)}\|_2\le 1$ and $\langle x^{(t)},g^{(t)}\rangle\le 0$: this holds trivially for padded steps, holds for private steps because $u^{(t)}$ averages unit rows satisfying $\langle\bar a_i,x^{(t)}\rangle\le 0$, and holds for separator steps because $\|p^{(t)}\|_2=1$ and $\langle p^{(t)},x^{(t)}\rangle\le 0$. Letting $v^{(t)}:=x^{(t)}+g^{(t)}$, it follows that $\|v^{(t)}\|_2^2\le\|x^{(t)}\|_2^2+1$, and hence
\begin{align*}
\|x^{(t+1)}\|_2^2
&= \|v^{(t)}+\zeta^{(t)}\|_2^2
\le \|x^{(t)}\|_2^2 + 1 + 2\langle v^{(t)},\zeta^{(t)}\rangle + \|\zeta^{(t)}\|_2^2.
\end{align*}

Let $\kappa:=2+4\log(1/p)$, so that $\kappa d\sigma^2\le 1$. Conditioned on the history prior to the epoch, $(\zeta^{(1)},\ldots,\zeta^{(T)})\sim\mathcal N(0,\sigma^2 I_{dT})$, so \Cref{lem:gaussian-quadratic} in dimension $dT$ guarantees $\sum_{t=1}^T\|\zeta^{(t)}\|_2^2\le\kappa dT\sigma^2\le T$ with probability at least $1-p$. Next, let $\mathbf{1}_t$ be the indicator of the event $\{\|x^{(s)}\|_2^2\le 5T \text{ for all } s\le t\}$, and let $Y:=\sum_{t=1}^T\mathbf{1}_t\langle v^{(t)},\zeta^{(t)}\rangle$. Conditioned on the filtration prior to sampling $\zeta^{(t)}$, the term $\mathbf{1}_t\langle v^{(t)},\zeta^{(t)}\rangle$ is zero-mean Gaussian with variance $\mathbf{1}_t\sigma^2\|v^{(t)}\|_2^2\le 6T\sigma^2$. Iterated conditioning yields $\mathbb E[e^{\lambda Y}]\le e^{3\lambda^2\sigma^2 T^2}$ for all $\lambda>0$, so optimizing $\lambda$ in Markov's inequality gives, with probability at least $1-p$,
\begin{align*}
Y
&\le \sqrt{12\sigma^2 T^2 \log(1/p)}
\le \sqrt{3\kappa\sigma^2}\,T
\le \sqrt{\frac{3}{d}}\,T
\le \sqrt{\frac{3}{2}}\,T,
\end{align*}
where the second inequality uses $4\log(1/p)\le\kappa$, the third uses $\kappa d\sigma^2\le 1$, and the fourth uses $d\ge 2$.

Suppose both high-probability bounds hold, which occurs with probability at least $1-2p$. We claim that $\|x^{(t)}\|_2^2\le 5T$ for all $t\in[T+1]$. Otherwise, let $t_1\in\{2,\ldots,T+1\}$ be the first index with $\|x^{(t_1)}\|_2^2>5T$. Then $\mathbf{1}_t=1$ for all $t<t_1$ and $\mathbf{1}_t=0$ for all $t\ge t_1$, so summing the one-step recursion over $t=1,\ldots,t_1-1$ gives
\begin{align*}
\|x^{(t_1)}\|_2^2
&\le (t_1-1) + \sum_{t=1}^{t_1-1}\|\zeta^{(t)}\|_2^2 + 2\sum_{t=1}^{t_1-1}\langle v^{(t)},\zeta^{(t)}\rangle \\
&\le T + \sum_{t=1}^T\|\zeta^{(t)}\|_2^2 + 2Y \\
&\le (2+\sqrt{6})\,T
< 5T,
\end{align*}
where the second inequality uses $t_1-1\le T$ and $\sum_{t<t_1}\langle v^{(t)},\zeta^{(t)}\rangle=Y$. This contradiction proves $\|x^{(T+1)}\|_2^2\le 5T$, and therefore $\|\widehat\mu\|_2=\|x^{(T+1)}\|_2/T\le\sqrt{5/T}$.
\end{proof}

\begin{lemma}[Noisy stopping; cf.\ {\cite[Lemma~11]{ene2025solving}}]
\label{fact:utility-stopping}
For any parameters $\nu\ge 0$, $\varepsilon_1,\delta_1,\beta_1\in(0,1)$, and integers $\tau,T,R_{\mathrm{try}}\ge 1$, let
\begin{align*}
\Gamma
&:= \nu + \frac{1}{\varepsilon_1}\Bigl(\log\frac{2TR_{\mathrm{try}}}{\delta_1} + \log\frac{\tau T}{\beta_1}\Bigr).
\end{align*}
In an execution of~\cref{alg:private-perceptron} over at most $\tau$ epochs, with probability at least $1-\beta_1$, any solution $x^{(t)}$ returned by~\cref{alg:private-perceptron-epoch} violates at most $\Gamma$ rows of its epoch input $A$.
\end{lemma}

\begin{proof}
Across the entire run, at most $\tau T$ private steps draw independent Laplace variables $Y\sim\Lap(1/\varepsilon_1)$ to form $\widehat N^{(t)}=N^{(t)}+Y-\varepsilon_1^{-1}\log(2TR_{\mathrm{try}}/\delta_1)$. Each draw satisfies $\Pr\bigl[Y<-\varepsilon_1^{-1}\log(\tau T/\beta_1)\bigr]=\beta_1/(2\tau T)$. Because the stopping time is data- and noise-dependent, we take a union bound over all at most $\tau T$ draws, showing that $Y\ge -\varepsilon_1^{-1}\log(\tau T/\beta_1)$ holds simultaneously for every private step with probability at least $1-\beta_1/2$. On this event, returning via $N^{(t)}=0$ incurs zero violations, whereas returning via $\widehat N^{(t)}\le\nu$ implies
\begin{align*}
N^{(t)}
&= \widehat N^{(t)} - Y + \frac{1}{\varepsilon_1}\log\frac{2TR_{\mathrm{try}}}{\delta_1}
\le \nu + \frac{1}{\varepsilon_1}\log\frac{\tau T}{\beta_1} + \frac{1}{\varepsilon_1}\log\frac{2TR_{\mathrm{try}}}{\delta_1}
= \Gamma. 
\end{align*}
\end{proof}

\begin{proposition}[Positive-margin mixed solver]
\label{prop:utility-mixed}
Assume $\rho(A,\mathcal K)\ge 2\rho_0>0$ with $\rho_0\le 1/2$, and let $\varepsilon,\delta,\beta\in(0,1)$ with $\beta_1=\beta/4$. \cref{alg:private-perceptron} is $(\varepsilon,\delta)$-DP and always outputs either $\bot$ or a point in $\mathcal K$. Moreover, with probability at least $1-\beta$, it outputs a point $x\in\mathcal K$ violating at most $\Gamma$ private rows, where $\Gamma$ is defined in \Cref{fact:utility-stopping} under the parameter choices of~\cref{alg:private-perceptron} and satisfies
\begin{align*}
\Gamma
&= \widetilde O\!\left(\frac{d^{3/2}}{\varepsilon}\sqrt{\log\frac{1}{\rho_0}}\right),
\end{align*}
where $\widetilde{O}$ hides polylogarithmic factors in $d$, $1/\delta$, $1/\beta$, and $\log(1/\rho_0)$.
\end{proposition}

\begin{proof}
For each epoch $\ell$, let $F_\ell$ and $P_\ell:=F_\ell\cap\mathbb B^d$ denote its mixed feasible cone and unit-ball section. Because~\cref{alg:private-perceptron-epoch} returns a candidate solution $z$ only when $z\in B^{-1}\mathcal K$, the mapped point $Bz$ returned by~\cref{alg:private-perceptron} always lies in $\mathcal K$. Likewise, since $A$ and $B$ undergo identical right-multiplications by $M^{-1}$, the rows active in epoch $\ell$ are positive scalar multiples of the input rows right-multiplied by $B$, so $Bz$ violates precisely the input rows whose transformed counterparts are violated at $z$.

\emph{Failure events.} Let $\tau_0:=1/(16\sqrt{\pi d})$, $p:=\beta_1/(4\tau R_{\mathrm{try}})$, $\kappa:=2+4\log(1/p)$, and $L:=T\tau/(\beta_1\delta_1)$. Under the parameter choices of~\cref{alg:private-perceptron} with a sufficiently large implied constant in $\nu$, we have $\theta^2=O\bigl(1/(d^2\log^2 L)\bigr)$, $\sigma^2=O\bigl(1/(d\log^2 L)\bigr)$, and $\kappa=O(\log L)$, which give
\begin{align*}
\kappa\theta^2 &\le \frac{\tau_0^2}{20d}
\qquad\text{and}\qquad
\kappa d\sigma^2 \le 1.
\end{align*}
For epoch $\ell$, let $\Sigma_\ell:=\Sigma_{F_\ell}$ be the covariance matrix from \Cref{lem:volume-increase}, which is measurable with respect to the history prior to epoch $\ell$, and let $\xi_\ell:=\sum_{t=1}^T\zeta^{(t)}$, conceptually sampling the remaining Gaussian vectors even if the perceptron or rescaling loop terminates early. Conditioned on the history before epoch $\ell$, we have $\xi_\ell\sim\mathcal N(0,T\sigma^2 I)$, while each $\gamma^{(s)}\sim\mathcal N(0,\theta^2 I)$ is independent of all preceding randomness. Define four bad events:
\begin{itemize}
\item ${\cal E}_1$: some epoch reaches its rescaling phase with $\|\widehat\mu\|_2>\sqrt{5/T}$;
\item ${\cal E}_2$: some trial in an epoch $\ell$ satisfies $\|\gamma^{(s)}\|_2^2>\kappa d\theta^2$ or $(\gamma^{(s)})^\top\Sigma_\ell\gamma^{(s)}>\kappa\theta^2$, or some epoch $\ell$ satisfies $\xi_\ell^\top\Sigma_\ell\xi_\ell>\kappa T\sigma^2$;
\item ${\cal E}_3$: some epoch reaches its rescaling phase with $\|c^{(s)}\|_2<2\tau_0$ across all $R_{\mathrm{try}}$ trials;
\item ${\cal E}_4$: the failure event of \Cref{fact:utility-stopping}.
\end{itemize}
Since $\kappa d\sigma^2\le 1$, applying \Cref{lem:support-function-bound} and a union bound over the at most $\tau$ epochs gives $\Pr[{\cal E}_1]\le 2\tau p\le\beta_1$. Applying \Cref{lem:gaussian-quadratic} to each of the at most $2\tau R_{\mathrm{try}}+\tau\le 4\tau R_{\mathrm{try}}$ quadratic forms in ${\cal E}_2$ gives $\Pr[{\cal E}_2]\le\beta_1$. Conditioned on the perceptron phase of an epoch, the $R_{\mathrm{try}}$ trials are independent and, by \Cref{lem:gaussian-subset,lem:support-function-bound}, each satisfies $\|c^{(s)}\|_2\ge 2\tau_0$ with probability at least $1/20$; hence all $R_{\mathrm{try}}\ge 1000\log(\tau/\beta_1)$ trials fail with probability at most $(19/20)^{R_{\mathrm{try}}}\le\beta_1/\tau$, giving $\Pr[{\cal E}_3]\le\beta_1$ by a union bound. Finally, $\Pr[{\cal E}_4]\le\beta_1$ by \Cref{fact:utility-stopping}. Because $\beta_1=\beta/4$, none of ${\cal E}_1,\ldots,{\cal E}_4$ occurs with probability at least $1-\beta$; assume henceforth that all four complements hold.

\emph{Volume increase.} Suppose epoch $\ell$ reaches its rescaling phase, and let $\mu:=\sum_j\widetilde\lambda_j\widetilde a_j$ and $\bar\xi:=\xi_\ell/T$, so that $\widehat\mu=\mu+\bar\xi$ by \Cref{lem:support-function-bound}. Outside ${\cal E}_1\cup {\cal E}_2\cup {\cal E}_3$, at least one trial $s$ satisfies $\|c^{(s)}\|_2\ge 2\tau_0$, while $\|\gamma^{(s)}\|_2\le\sqrt{\kappa d}\,\theta\le\tau_0/\sqrt{20}<\tau_0$ and $T\ge(768\sqrt{\pi})^2 d$ give
\begin{align*}
\|\widehat\mu\|_2\le \sqrt{\frac{5}{T}}
\le \frac{\tau_0}{4},
\qquad 
\|\widehat c^{(s)}\|_2
\ge \|c^{(s)}\|_2 - \|\gamma^{(s)}\|_2
\ge 2\tau_0 - \tau_0
= \tau_0
\ge 4\|\widehat\mu\|_2.
\end{align*}
Thus trial $s$ passes both acceptance tests, so the epoch returns $(\widehat c^{(s^*)},\widehat\mu)$ for the first passing trial $s^*$. Moreover, $F_\ell$ has nonempty interior because $P_1$ contains a Euclidean ball by \Cref{lem:initial-volume} and each rescaling matrix is invertible, while \Cref{lem:support-function-bound} implies that every $\widetilde a_j$ is nonnegative on $F_\ell$ and hence $0\le\langle c^{(s^*)},x\rangle\le\langle\mu,x\rangle$ for all $x\in F_\ell$ since $c^{(s^*)}$ is a sub-sum of $\mu$. Using $\|\widehat c^{(s^*)}\|_2\ge\tau_0$, the complement of ${\cal E}_2$, and $\kappa\sigma^2\le 1/d$ together with $T\ge(768\sqrt{\pi})^2 d$, we bound the exponent in \Cref{lem:volume-increase} by
\begin{align*}
\frac{d\,\bigl((\gamma^{(s^*)})^\top\Sigma_\ell\gamma^{(s^*)} + 2\bar\xi^\top\Sigma_\ell\bar\xi\bigr)}{\|\widehat c^{(s^*)}\|_2^2}
&\le \frac{d}{\tau_0^2}\Bigl(\kappa\theta^2 + \frac{2\kappa\sigma^2}{T}\Bigr)
\le \frac{d}{\tau_0^2}\Bigl(\frac{\tau_0^2}{20d} + \frac{\tau_0^2}{20d}\Bigr)
= \frac{1}{10}.
\end{align*}
Since updating $A$ and $B$ replaces $F_\ell$ with $MF_\ell$, \Cref{lem:volume-increase} yields
\begin{align*}
\mathrm{vol}(P_{\ell+1})
&\ge \sqrt{\frac{3}{2}}\,e^{-1/10}\,\mathrm{vol}(P_\ell)
\ge 1.1\,\mathrm{vol}(P_\ell).
\end{align*}

\emph{Termination and bound on $\Gamma$.} Combining $\mathrm{vol}(P_1)\ge\rho_0^d\,\mathrm{vol}(\mathbb B^d)$ from \Cref{lem:initial-volume} with $\mathrm{vol}(P_\ell)\le\mathrm{vol}(\mathbb B^d)$ shows that at most $\lfloor d\log(1/\rho_0)/\log 1.1\rfloor=\tau-1$ rescalings can occur. Thus, at least one of the $\tau$ epochs halts in its perceptron phase and, outside ${\cal E}_4$, returns a point in $\mathcal K$ violating at most $\Gamma$ private rows. Under the parameter choices in~\cref{alg:private-perceptron}, we have $M_p=\Theta\bigl(d^2\log(1/\rho_0)\bigr)$ and $M_r=\Theta\bigl(d\log(1/\rho_0)\log(\tau/\beta_1)\bigr)$, so
\begin{align*}
Q
&= 4M_p + dM_r
= \Theta\!\left(d^2\log\frac{1}{\rho_0}\cdot\log\frac{\tau}{\beta_1}\right),
&
\log\frac{1}{\delta_1}
&= \log\frac{2M_0}{\delta}
= O\!\left(\log\frac{d\log(1/\rho_0)\log(1/\beta)}{\delta}\right),
\end{align*}
where the first estimate uses $\log(\tau/\beta_1)>1$ since $\beta_1=\beta/4<1/4$. Substituting $\varepsilon_1=\varepsilon/\bigl(8\sqrt{Q\log(2/\delta)}\bigr)$ and our bound on $\log(1/\delta_1)$ into $\nu$ and $\Gamma$ in \Cref{fact:utility-stopping} gives $\Gamma=\widetilde O\bigl((d^{3/2}/\varepsilon)\sqrt{\log(1/\rho_0)}\bigr)$.

The proof of the proposition then completes.
\end{proof}

\section{Proof of the Main Theorem}
\label{app:upper-proof}

In this appendix, we reduce any $R$-feasible SCP instance to a mixed homogeneous conic system with positive margin and prove \Cref{thm:main}. Specifically, relaxing each private row by $\eta$ and homogenizing yields a system over the public cone $C_K=K\times\R$ with mixed margin $\Omega\bigl(\eta/(UR\sqrt{r})\bigr)$, as shown in \Cref{thm:margin}, which is then combined with \Cref{prop:utility-mixed} to establish \Cref{thm:main}. For comparison, \Cref{subsec:song-tuning} derives the violation count of the covering-SCP framework of~\cite{song2025differential} at additive slack $\eta$.

\subsection{Review of Euclidean Jordan algebras}
\label{subsec:eja-background}

We briefly recall the definitions underlying properties (J1)--(J4) in \Cref{subsec:ejas} (see~\cite{faraut1994analysis} for proofs). A \emph{Euclidean Jordan algebra} (EJA) $J$ is a finite-dimensional real inner-product space equipped with a commutative bilinear product $\circ$ and an \emph{identity element} $e\in J$ satisfying $e\circ x=x$ for all $x\in J$, such that the Jordan identity $x^2\circ(x\circ y)=x\circ(x^2\circ y)$ holds with $x^2:=x\circ x$ and the inner product satisfies the associativity relation $\langle x\circ y,z\rangle=\langle x,y\circ z\rangle$. Every symmetric cone arises as the cone of squares $K=\{x\circ x:x\in J\}$ of such an algebra $J$, and we write $k:=\dim(J)$. An element $q\in J$ is an \emph{idempotent} if $q\circ q=q$, and is \emph{primitive} if it is nonzero and cannot be written as the sum of two nonzero idempotents $q',q''$ satisfying $q'\circ q''=0$. The \emph{rank} $r$ of $J$ and of $K$ is the number of eigenvalues, counted with multiplicity, of each element in the spectral decomposition of (J2).

Throughout the paper, we fix the canonical trace inner product $\langle x,y\rangle:=\mathrm{Tr}(x\circ y)$ of~\cite[Chapter~III]{faraut1994analysis}, under which every primitive idempotent has unit norm and (J1)--(J4) take their standard form. Identifying $J$ with $\R^k$ via an isometry for this inner product means that all Euclidean norms, including the bounds $U$ and $R$, coincide with trace norms. Concretely, for $\R_+^n$ this identification is the identity; for the PSD cone it vectorizes a symmetric matrix with off-diagonal entries scaled by $\sqrt{2}$, recovering the Frobenius norm; and for the Lorentz cone, whose trace inner product equals twice the standard Euclidean inner product, it scales the coordinates in \Cref{tab:cones} by $\sqrt{2}$, leaving the cone itself unchanged. In the PSD case, the Jordan product is $X\circ Y=(XY+YX)/2$, the identity is $e=I$, the trace inner product is $\langle X,Y\rangle=\mathrm{Tr}(XY)$, primitive idempotents are rank-one projectors $vv^\top$, and Jordan frames correspond to orthonormal eigenbases.

\subsection{Relaxation and homogenization give positive margin}
\label{subsec:margin-reduction}

Given private rows $(a_i,b_i)\in J\times\R$ for $i\in[m]$ and a slack parameter $\eta>0$, we write points of $J\times\R$ as $y=(x,t)$ and define the homogenized private rows by
\begin{align*}
m_i := (-a_i,\,b_i+\eta),
\qquad
\bar a_i := \frac{m_i}{\|m_i\|_2}.
\end{align*}
Whenever $y=(x,t)$ satisfies $t>0$ and $\langle m_i,y\rangle>0$, dividing by $t$ gives $\langle a_i,x/t\rangle<b_i+\eta$. If $m_i=0$, which occurs only when $a_i=0$ and $b_i=-\eta$, the normalization is undefined, and we drop any such row from the input since every $x\in J$ already satisfies $\langle a_i,x\rangle=0=b_i+\eta$. Hence, we can assume no row has $m_i=0$.

Because $x$ must lie in $K$, the homogenized variable $y=(x,t)$ ranges over the public cone
\begin{align*}
C_K := K\times\R.
\end{align*}
The cone $C_K$ is closed and convex, though it contains the line $\{0\}\times\R$ and is therefore neither pointed nor symmetric. Since $K$ is self-dual, the dual cone of $C_K$ is $C_K^*=K^*\times\{0\}=K\times\{0\}$. Both~\cref{alg:private-perceptron} and~\cref{prop:utility-mixed} require only that the public cone be closed and convex and admit a separation oracle: a point $(x,t)$ lies outside $C_K$ if and only if $x\notin K$, in which case $(q,0)$ serves as a unit separator whenever $q$ is a unit separator of $x$ from $K$.

Dehomogenizing $y=(x,t)$ to $x/t$ requires $t>0$, so we introduce the auxiliary row
\begin{align*}
e_t := (0,1)\in J\times\R.
\end{align*}
In the proof of \Cref{thm:main}, we append enough copies of $e_t$ to enforce $t>0$ with high probability. Let $A^+$ denote the matrix with rows $\bar a_1,\ldots,\bar a_m$ together with $e_t$. 

The next theorem shows that on any $R$-feasible instance, the mixed margin of $(A^+,C_K)$ scales at least linearly with $\eta$. Because the violation bound in \Cref{prop:utility-mixed} depends on the margin $\rho_0$ only through $\sqrt{\log(1/\rho_0)}$, our final violation bound depends only logarithmically on $1/\eta$.

\begin{theorem}[Positive margin after relaxation and homogenization]
\label{thm:margin}
Assume $\|a_i\|_2\le U$ and $|b_i|\le U$ for all $i\in[m]$. Suppose there exists an exactly feasible point $x^\star\in K$ satisfying $\langle a_i,x^\star\rangle\le b_i$ for all $i\in[m]$, with $\|x^\star\|_2\le R$ and $R\ge 1$. Then for every slack $\eta\in(0,U]$, the augmented system has mixed margin
\begin{align*}
\rho(A^+,C_K)
&\ge \frac{\eta}{2\sqrt{10}\,UR\sqrt{r}}.
\end{align*}
\end{theorem}

\begin{proof}
We construct a unit witness $y_\eta$ that strictly satisfies all constraints of the augmented system and lower-bound its inner product against each constraint family.

\emph{Construction of the witness.}
Let $e\in J$ be the identity element of the EJA, which satisfies $\|e\|_2=\sqrt{r}$ and $\lambda_{\min}(e)=1$, and let
\begin{align*}
e_K := \frac{R}{\sqrt{r}}\,e,
\qquad\text{so that}\qquad
\|e_K\|_2 = R
\quad\text{and}\quad
\lambda_{\min}(e_K) = \frac{R}{\sqrt{r}}.
\end{align*}
Next, let the mixing weight $\alpha$ and the interpolated point $x_\eta$ be given by
\begin{align*}
\alpha := \frac{\eta}{2U(R+1)},
\qquad
x_\eta := (1-\alpha)x^\star + \alpha e_K.
\end{align*}
Convexity of $K$ ensures $x_\eta\in K$, and the triangle inequality gives $\|x_\eta\|_2\le(1-\alpha)\|x^\star\|_2+\alpha\|e_K\|_2\le R$. By concavity of $\lambda_{\min}$ on $K$ together with $\lambda_{\min}(x^\star)\ge 0$, we obtain
\begin{align*}
\lambda_{\min}(x_\eta)
&\ge (1-\alpha)\lambda_{\min}(x^\star) + \alpha\lambda_{\min}(e_K)
\ge \frac{\alpha R}{\sqrt{r}}
= \frac{\eta R}{2U(R+1)\sqrt{r}}
\ge \frac{\eta}{4U\sqrt{r}},
\end{align*}
where the final inequality uses $R\ge 1$ to bound $R/(R+1)\ge 1/2$.

To lift $x_\eta$ to the homogenized space, let $\widetilde y_\eta:=(x_\eta,1)\in J\times\R$. Since $\|x_\eta\|_2\le R$ and $R\ge 1$, its norm $S:=\|\widetilde y_\eta\|_2$ satisfies
\begin{align*}
S
&= \sqrt{\|x_\eta\|_2^2+1}
\le \sqrt{R^2+1}
\le \sqrt{2}\,R.
\end{align*}
Let $y_\eta:=\widetilde y_\eta/S$ be the corresponding unit vector. We now lower-bound $\langle\cdot,y_\eta\rangle$ across all three constraint types.

\emph{Margin against private rows.}
Fix $i\in[m]$. Expanding $m_i=(-a_i,b_i+\eta)$ and $x_\eta=(1-\alpha)x^\star+\alpha e_K$ gives
\begin{align*}
\langle m_i,\widetilde y_\eta\rangle
&= (b_i+\eta) - \langle a_i,x_\eta\rangle
= \eta + (1-\alpha)\bigl(b_i-\langle a_i,x^\star\rangle\bigr) + \alpha\bigl(b_i-\langle a_i,e_K\rangle\bigr).
\end{align*}
Feasibility of $x^\star$ implies $b_i-\langle a_i,x^\star\rangle\ge 0$. For the remaining term, the Cauchy--Schwarz inequality gives $|\langle a_i,e_K\rangle|\le\|a_i\|_2\,\|e_K\|_2\le UR$, which combined with $|b_i|\le U$ yields $b_i-\langle a_i,e_K\rangle\ge -U(R+1)$. Substituting the definition of $\alpha$, we obtain
\begin{align*}
\langle m_i,\widetilde y_\eta\rangle
&\ge \eta - \alpha U(R+1)
= \eta - \frac{\eta}{2}
= \frac{\eta}{2}.
\end{align*}
Moreover, because $\|a_i\|_2\le U$, $|b_i|\le U$, and $\eta\le U$, the unnormalized row norm satisfies
\begin{align*}
\|m_i\|_2
&= \sqrt{\|a_i\|_2^2+(b_i+\eta)^2}
\le \sqrt{U^2+(2U)^2}
= \sqrt{5}\,U.
\end{align*}
Combining this bound with $S\le\sqrt{2}\,R$ and $r\ge 1$ gives
\begin{align*}
\langle\bar a_i,y_\eta\rangle
&= \frac{\langle m_i,\widetilde y_\eta\rangle}{\|m_i\|_2\,S}
\ge \frac{\eta/2}{\sqrt{5}\,U\cdot\sqrt{2}\,R}
= \frac{\eta}{2\sqrt{10}\,UR}
\ge \frac{\eta}{2\sqrt{10}\,UR\sqrt{r}}.
\end{align*}

\emph{Margin against the row $e_t$.}
For $e_t=(0,1)$, using $S\le\sqrt{2}\,R$ together with $\eta\le U$ and $r\ge 1$ yields
\begin{align*}
\langle e_t,y_\eta\rangle
&= \frac{1}{S}
\ge \frac{1}{\sqrt{2}\,R}
\ge \frac{\eta}{\sqrt{2}\,UR\sqrt{r}}
\ge \frac{\eta}{2\sqrt{10}\,UR\sqrt{r}}.
\end{align*}

\emph{Margin against the public cone.}
Let $h=(u,0)\in C_K^*=K\times\{0\}$ be any unit vector, so that $u\in K$ and $\|u\|_2=1$. Then $\langle h,y_\eta\rangle=\langle u,x_\eta\rangle/S$. Since $x_\eta-\lambda_{\min}(x_\eta)\,e\in K$ and $K$ is self-dual, we have $\langle u,x_\eta\rangle\ge\lambda_{\min}(x_\eta)\,\langle u,e\rangle$. Expanding $u=\sum_{j=1}^r\lambda_j q_j$ in a Jordan frame with $\lambda_j\ge 0$ and $\sum_{j=1}^r\lambda_j^2=1$ gives
\begin{align*}
\langle u,e\rangle
&= \sum_{j=1}^r \lambda_j
\ge \sqrt{\sum_{j=1}^r \lambda_j^2}
= 1.
\end{align*}
Combining $\langle u,x_\eta\rangle\ge\lambda_{\min}(x_\eta)$ with $\lambda_{\min}(x_\eta)\ge\eta/(4U\sqrt{r})$ and $S\le\sqrt{2}\,R$ gives
\begin{align*}
\langle h,y_\eta\rangle
&\ge \frac{\lambda_{\min}(x_\eta)}{S}
\ge \frac{\eta/(4U\sqrt{r})}{\sqrt{2}\,R}
= \frac{\eta}{4\sqrt{2}\,UR\sqrt{r}}
\ge \frac{\eta}{2\sqrt{10}\,UR\sqrt{r}},
\end{align*}
where the last inequality uses $4\sqrt{2}<2\sqrt{10}$. 

Taking the minimum over all three constraint families establishes $\rho(A^+,C_K)\ge\eta/(2\sqrt{10}\,UR\sqrt{r})$, completing the proof of the theorem.
\end{proof}


We now state and prove the main theorem for our upper bound.
\begin{theorem}[Row-private relaxed symmetric cone feasibility, formal version of~\cref{thm:upper-informal}]
\label{thm:main}
Let $J$ be a Euclidean Jordan algebra of dimension $k$ and rank $r$ with symmetric cone $K\subseteq J$, and let $U>0$, $R\ge 1$, $\eta\in(0,U]$, and $\varepsilon,\delta,\beta\in(0,1)$. There is an efficient $(\varepsilon,\delta)$-differentially private algorithm under add/remove row adjacency that takes a list of rows $(a_i,b_i)\in J\times\R$ for $i\in[m]$ and always outputs a point $\widehat x\in K$. 

Moreover, on every $R$-feasible instance of \Cref{def:problem}, where $\|a_i\|_2\le U$ and $|b_i|\le U$ for all $i\in[m]$ and there exists $x^\star\in K$ with $\|x^\star\|_2\le R$ satisfying $\langle a_i,x^\star\rangle\le b_i$ for all $i\in[m]$, the output satisfies with probability at least $1-\beta$
\begin{align*}
\bigl|\{i\in[m]:\langle a_i,\widehat x\rangle>b_i+\eta\}\bigr|
&\le \widetilde O\!\left(
\frac{k^{3/2}}{\varepsilon}
\sqrt{\log\Bigl(2+\frac{UR\sqrt{r}}{\eta}\Bigr)}
\right),
\end{align*}
where $\widetilde O(\cdot)$ hides factors polylogarithmic in $k$, $1/\delta$, $1/\beta$, and $\log(2+UR\sqrt{r}/\eta)$.
\end{theorem}

\begin{proof}
\emph{Algorithm.} Working in ambient dimension $d=k+1$, we form the normalized homogenized rows $\bar a_i=m_i/\|m_i\|_2$ from \Cref{subsec:margin-reduction} with $m_i=(-a_i,b_i+\eta)$, dropping any row with $m_i=0$, together with the auxiliary row $e_t=(0,1)$ and the public cone $C_K=K\times\R$. Let
\begin{align*}
\rho_0 := \frac{\eta}{4\sqrt{10}\,UR\sqrt{r}},
\end{align*}
which depends only on public parameters and satisfies $\rho_0\le 1/(4\sqrt{10})<1/2$ since $\eta\le U$, $R\ge 1$, and $r\ge 1$. Let $\Gamma$ be the violation threshold from~\cref{fact:utility-stopping} evaluated at dimension $d=k+1$, margin $\rho_0$, privacy parameters $(\varepsilon,\delta)$, failure probability $\beta$, and $\beta_1=\beta/4$ as in~\cref{alg:private-perceptron}. Both $\Gamma$ and $B_t:=\lceil\Gamma\rceil+1$ depend only on public quantities. We run~\cref{alg:private-perceptron} with public cone $C_K$ on the row list consisting of $\bar a_1,\ldots,\bar a_m$ followed by $B_t$ identical copies of $e_t$. If the run returns $y=(x,t)$ with $t>0$, we output $\widehat x:=x/t$; if it returns $\bot$ or a point with $t\le 0$, we output $\widehat x:=0$. By \Cref{prop:utility-mixed}, \cref{alg:private-perceptron} returns only points in $C_K$, so $x\in K$ and hence $\widehat x\in K$ on every execution.

\emph{Privacy.} Dropping rows with $m_i=0$ and normalizing act independently on each private row, while the $B_t$ appended copies of $e_t$ are input-independent, so add/remove neighboring inputs produce neighboring or identical row lists. By~\cref{prop:privacy-composition}, the call to~\cref{alg:private-perceptron} is $(\varepsilon,\delta)$-DP, and post-processing via \Cref{fact:post-processing} preserves $(\varepsilon,\delta)$-DP for $\widehat x$.

\emph{Utility.} Consider any $R$-feasible instance. By \Cref{thm:margin}, duplicating $e_t$ leaves the minimum row margin unchanged, so
\begin{gather*}
\rho(A^+,C_K)\ge \frac{\eta}{2\sqrt{10}\,UR\sqrt{r}}
= 2\rho_0,
\\
\log\frac{1}{\rho_0}= \log(4\sqrt{10}) + \log\frac{UR\sqrt{r}}{\eta}
= O\!\left(\log\Bigl(2+\frac{UR\sqrt{r}}{\eta}\Bigr)\right),
\end{gather*}
where the second line uses $UR\sqrt{r}/\eta\ge 1$. Because each of the four failure events in the proof of~\cref{prop:utility-mixed} occurs with probability at most $\beta_1=\beta/4$, \cref{prop:utility-mixed} implies that with probability at least $1-\beta$, the run returns a point $y=(x,t)\in C_K$ violating at most $\Gamma$ rows, where a row is violated whenever its inner product with $y$ is nonpositive. Since all $B_t$ copies of $e_t$ have inner product $t$ with $y$, having $t\le 0$ would incur $B_t>\Gamma$ violations. Thus $t>0$ and $\widehat x=x/t$ on this success event. For every kept row $\bar a_i$ that is not violated by $y$, we have
\begin{align*}
\langle m_i,(x,t)\rangle > 0
\quad\Longrightarrow\quad
-\langle a_i,x\rangle + t(b_i+\eta) > 0
\quad\Longrightarrow\quad
\langle a_i,\widehat x\rangle < b_i+\eta,
\end{align*}
while any dropped row with $m_i=0$ is automatically $\eta$-satisfied. Hence the number of private rows with $\langle a_i,\widehat x\rangle>b_i+\eta$ is at most $\Gamma$. Substituting $d=k+1$ and our bound on $\log(1/\rho_0)$ into $\Gamma=\widetilde O\bigl(d^{3/2}\sqrt{\log(1/\rho_0)}/\varepsilon\bigr)$ from~\cref{prop:utility-mixed} completes the proof.
\end{proof}

\subsection{Comparison with~\texorpdfstring{\cite{song2025differential}}{Song et al.}}
\label{subsec:song-tuning}

We provide a quantitative comparison of \Cref{thm:main} with the row-private covering-SCP solver of~\cite{song2025differential} when its additive error parameter equals the slack $\eta$.\footnote{Theorem, lemma, and remark numbers refer to version~2 of their arXiv preprint (arXiv:2509.16915v2).}

\paragraph{The result of~\cite{song2025differential}.} A covering SCP over a simple EJA of rank $r$ minimizes $\mathrm{Tr}(x)$ over $x\in K$ subject to $\langle a_i,x\rangle\ge 1$ for $i\in[m]$, where each $a_i\in K$ satisfies $\lambda_{\max}(a_i)\le 1$. Because 
$$
\langle a_i,x\rangle\le\lambda_{\max}(a_i)\,\mathrm{Tr}(x)\le\mathrm{Tr}(x)\qquad \forall x\in K,
$$ 
its optimal value satisfies $\mathrm{OPT}\ge 1$. As noted in their Remarks~B.8 and~B.12, \cite[Theorem~B.15]{song2025differential} requires both $\mathrm{OPT}$ and an upper bound $\bar m\ge m+1$ on the row count as public inputs. For any $\alpha\in(0,\mathrm{OPT}]$, their $(\varepsilon,\delta)$-DP solver outputs $x_{\mathrm{S}}\in K$ with $\mathrm{Tr}(x_{\mathrm{S}})=\mathrm{OPT}$ such that, with probability at least $1-\beta$, fewer than $s$ rows satisfy $\langle a_i,x_{\mathrm{S}}\rangle<1-\alpha$, provided
\begin{align*}
m\geq s
&\ge \Omega\left(\frac{\mathrm{OPT}^2}{\varepsilon\,\alpha^2}\sqrt{\log(1/\delta)\log\bar m}\,\Bigl(k_C\log\frac{2\,\mathrm{OPT}}{\alpha}+\log\frac{\log\bar m}{\beta}\Bigr)\right).
\end{align*}
By their Lemma~B.13, $k_C=d_P(r-1)+1$ denotes the dimension of the cone of scaled primitive idempotents, where $d_P$ is the Peirce constant of $J$: specifically, $k_C=r$ for $r\times r$ real symmetric matrices, $k_C\le 4r$ for complex and quaternionic Hermitian matrices, $k_C=17$ for the Albert algebra, and $k_C=k-1$ for the $k$-dimensional spin factor associated with the Lorentz cone. Since $\langle a_i,x\rangle\ge 0$ for all $x\in K$, this guarantee is nontrivial only when $\alpha<1$. We therefore assume $\eta<1$ and let $\alpha=\eta$, which yields 
\begin{align}\label{eq:app_song_bound}
    \widetilde O\bigl(k_C\,\mathrm{OPT}^2\sqrt{\log\bar m}/(\varepsilon\eta^2)\bigr)
\end{align}
violated rows, hiding polylogarithmic factors in $1/\delta$, $1/\beta$, $\log\bar m$, and $\mathrm{OPT}/\eta$.

\paragraph{Comparison.} Rewriting each covering constraint as $\langle -a_i,x\rangle\le -1$ places the problem in the form of \Cref{def:problem}. Because every $a_i\in K$ has eigenvalues in $[0,1]$, while any optimal solution $x^\star\in K$ satisfies $\|x^\star\|_2\le\mathrm{Tr}(x^\star)=\mathrm{OPT}$, a covering SCP is an $R$-feasible instance with parameters
\begin{align*}
U = \sqrt{r},
\qquad
R = \mathrm{OPT},
\qquad
\frac{UR\sqrt{r}}{\eta} = \frac{r\,\mathrm{OPT}}{\eta}.
\end{align*}
Thus $\mathrm{OPT}/\eta$ matches our scale-to-slack ratio up to a factor of $r$, and \Cref{thm:main} yields $$\widetilde O\bigl(k^{3/2}\sqrt{\log(2+r\,\mathrm{OPT}/\eta)}/\varepsilon\bigr)$$ violated rows with no dependence on $m$. Moreover, appending public copies of the objective constraint $\langle e/\mathrm{OPT},x\rangle\le 1$, as discussed in \Cref{sec:limitations}, ensures $\mathrm{Tr}(\widehat x)\le(1+\eta)\,\mathrm{OPT}$ as well.

Up to logarithmic factors, our bound improves upon~\eqref{eq:app_song_bound} whenever $k_C\,\mathrm{OPT}^2/\eta^2\ge k^{3/2}$: for $r\times r$ real symmetric matrices, where $k_C=\Theta(r)$ and $k=\Theta(r^2)$, this holds once $\mathrm{OPT}/\eta=\Omega(r)$, and for the Lorentz cone, where $k_C=k-1$, it holds once $\mathrm{OPT}/\eta=\Omega(k^{1/4})$. Finally, the linear rank factor $k_C=r$ for covering SDPs does not contradict our $\Omega(k)=\Omega(r^2)$ lower bound in \Cref{cor:lower-lorentz-psd}, because the hard instances constructed there include the upper-bound constraints \eqref{eq:lower-upper-row} of the form $\langle UF_i,z\rangle\le U\mu_{w_i}$ and are therefore not covering programs.

\section{Obstruction to Perturb-and-Eliminate for Exact SCP Feasibility}
\label{app:hardness-exact}

In this appendix, we substantiate \Cref{rmk:upper-slack} and \Cref{sec:limitations} and explain why our formulation allows additive slack on the private rows. For linear programs with integer data, the row-private solvers of~\cite{kaplan2024differentially,ene2025solving} violate a bounded number of rows while satisfying all remaining rows exactly. Both algorithms follow a perturb-and-eliminate strategy: they relax every row by a uniform perturbation $\gamma>0$ to create a positive margin, solve the perturbed system privately, and identify the original rows violated by the perturbed solution. When $\gamma$ is sufficiently small, every such row must hold with equality at some exact solution, allowing the algorithm to turn those rows into equalities and project onto a lower-dimensional subspace. Below, we show that for symmetric cone programs, this guarantee can fail unless $\gamma$ is doubly exponentially small in the dimension.

For an LP with integer data of magnitude at most $U$, Cramer's rule implies that every nonzero coordinate of a basic feasible solution has magnitude at least $\bigl((d+1)U\bigr)^{-(d+1)}$, where $d=k+1$ is the homogenized dimension in \Cref{sec:main-results}. Choosing $\gamma$ below this threshold separates inactive constraints from tight ones while keeping $\log(1/\gamma)=O\bigl(d\log(dU)\bigr)$, so the perceptron epoch bound $O\bigl(d\log(1/\rho_0)\bigr)$ remains polynomial in $d$ and $\log U$. No comparable bound holds for SCPs, as demonstrated by the following adaptation of Khachiyan's iterated-squaring construction~\cite{pataki2024exponential}.

\paragraph{An iterated-squaring SCP instance.}
Fix a chain length $n\ge 1$, let $t_0:=\tfrac{1}{2}$, and consider the system in variables $(t_1,\ldots,t_n,x,y)$ given by
\begin{align*}
t_i \ge 0,\quad t_i \ge t_{i-1}^2 \quad \text{for } i\in[n],
\qquad
x,y \ge 0,
\qquad
x+y = t_n.
\end{align*}
Every exact solution satisfies $t_n\ge 2^{-2^n}$, and setting $t_i=2^{-2^i}$ for $i\in[n]$ attains this lower bound.

To express this system as an instance of \Cref{def:problem}, let $k:=3n+3$ and take the public cone $K:=\mathcal Q^n\times\mathcal L^3\subseteq\R^k$, where $$\mathcal L^3:=\{(s,x,y)\in\R^3:s\ge\|(x,y)\|_2\}$$ is the three-dimensional Lorentz cone and $$\mathcal Q:=\{(u,v,w)\in\R^3:2uv\ge w^2,\ u,v\ge 0\}$$ is the rotated second-order cone. Because $\mathcal Q$ is an orthogonal image of $\mathcal L^3$~\cite{benTalNemirovski2001}, $K$ is a symmetric cone. For each $i\in[n]$, the $i$-th block constrains $(u_i,t_i,w_i)\in\mathcal Q$, while the final block constrains $(s,x,y)\in\mathcal L^3$ and therefore places no sign restriction on $x$ or $y$. As private rows, we include
\begin{align*}
2u_i &= 1 \quad\forall i\in[n], &
2w_1 &= 1, &
w_i &= t_{i-1} \quad\forall 2\le i\le n, \\
x+y &= t_n, &
-x &\le 0, &
-y &\le 0,
\end{align*}
writing each equality as a pair of opposite inequalities and duplicating each row more times than the violation budget of the solver. Every row satisfies $\|a\|_2\le 2$ and $|b|\le 1$. When all private rows hold without slack, they force $u_i=\tfrac{1}{2}$ and $w_i=t_{i-1}$ for all $i\in[n]$, reducing the $i$-th rotated cone constraint $2u_i t_i\ge w_i^2$ to $t_i\ge t_{i-1}^2$ with $t_i\ge 0$. Thus, exact satisfaction of the private rows over $K$ reproduces the iterated-squaring system above.

\paragraph{Failure of elimination unless $\gamma$ is doubly exponentially small.}
Now relax every private row by a uniform perturbation $\gamma>0$, let $\tau_n:=2^{-2^n}$, and suppose $\gamma\ge 2\tau_n$. Consider the candidate point in $K$ defined by
\begin{align*}
u_i := \tfrac{1}{2},
\qquad
t_i := 2^{-2^i},
\qquad
w_i := t_{i-1} \quad\forall i\in[n],
\qquad
x = y := -\frac{\tau_n}{4},
\qquad
s := \frac{\sqrt{2}\,\tau_n}{4}.
\end{align*}
Here, $(u_i,t_i,w_i)\in\mathcal Q$ holds with equality for each $i\in[n]$, and $(s,x,y)\in\mathcal L^3$ holds because $\|(x,y)\|_2=s$. At this point, $t_n=\tau_n$ and every private row is satisfied without slack except three:
\begin{align*}
-x &= \frac{\tau_n}{4} > 0, &
-y &= \frac{\tau_n}{4} > 0, &
t_n - x - y &= \frac{3\tau_n}{2} > 0,
\end{align*}
where the third inequality is one half of the split equality $x+y=t_n$, while the opposite half $x+y-t_n=-3\tau_n/2\le 0$ holds strictly. Because $\gamma\ge 2\tau_n$, all three violations lie in $(0,\gamma]$, so this point satisfies the $\gamma$-perturbed system completely.

In the elimination step of~\cite[Algorithm~3]{ene2025solving}, the rows converted into equalities are precisely those violated in the original system by a solution of the $\gamma$-perturbed system. For integer LPs with sufficiently small $\gamma$, \cite[Lemma~24]{ene2025solving} guarantees that the original system remains feasible after enforcing equality on these rows. On our SCP instance, however, enforcing equality on the three rows above forces $x=y=0$ and $t_n=x+y=0$, contradicting $t_n\ge\tau_n>0$ and making the reduced system infeasible.

Therefore, the invariant underlying the elimination step fails whenever $\gamma\ge 2^{1-2^n}$. For the LP analysis to apply to this instance, one must take $\gamma<2^{1-2^n}=2^{-2^{\Omega(k)}}$, which is doubly exponentially small in the ambient dimension $k$. Since the unperturbed system has zero margin, a perturbation of size $\gamma$ creates a margin $\rho_0$ of order at most $\gamma$, causing $\log(1/\rho_0)$ and the resulting epoch bound $O\bigl(d\log(1/\rho_0)\bigr)$ to grow exponentially in $k$. This obstruction motivates our relaxation in \Cref{def:problem}, which allows additive slack $\eta>0$ on the private rows while maintaining exact membership in $K$.

This construction rules out extending the perturb-and-eliminate analysis directly from LPs to SCPs. Whether a different polynomial-time private algorithm can achieve exact satisfaction on non-polyhedral SCPs with zero margin remains open, as discussed in \Cref{sec:limitations}.

\section{Lower Bound under Approximate Satisfaction}
\label{app:lower-bound}

For an instance $D$ consisting of private rows $\langle a_i,z\rangle\le b_i$, let
\begin{align*}
\operatorname{viol}_\eta(z;D)
&:= \bigl|\{i:\langle a_i,z\rangle>b_i+\eta\}\bigr|
\end{align*}
denote the number of $\eta$-violated constraints at $z$. In this appendix, we prove that for parameters in the regime of \Cref{cor:lower-add-remove}, every $(\varepsilon,\delta)$-DP algorithm under add/remove row adjacency must incur, with constant probability on some $R$-feasible instance,
\begin{align*}
\operatorname{viol}_\eta
&= \Omega\!\left(\frac{k}{\varepsilon}\log\min\!\left\{\frac{UR}{k\eta},\,\frac{\varepsilon}{\delta}\right\}\right).
\end{align*}
This bound already holds when the public cone is the nonnegative orthant $\R_+^k$, and extends with the same asymptotics (up to constant factors) to a single Lorentz cone or a positive semidefinite (PSD) cone of ambient dimension $k$.

Our proof combines a coordinate-wise packing construction with group privacy. In the hard instances constructed in \Cref{subsec:lower-hard-family}, each direction of an \emph{encoding frame} inside the cone carries an independent hidden symbol, enforced by a pair of opposite linear constraints. As we show in \Cref{subsec:lower-decoding}, $\eta$-satisfying both constraints along a frame direction pins down its coordinate tightly enough to recover that symbol uniquely. Therefore, violating too few rows forces the output to recover more than half of the hidden symbols, which cannot happen due to the group privacy shown in \Cref{subsec:lower-proof}. Because the geometry of the public cone enters only through the existence of an encoding frame, we first instantiate the frame for $\R_+^k$ in \Cref{subsec:lower-hard-family} and then construct frames of size $k-1$ for the Lorentz and PSD cones in \Cref{subsec:lower-nonpolyhedral}.

\subsection{Encoding frames and hard instances}
\label{subsec:lower-hard-family}

Fix $U>0$, dimension $k\ge 2$, radius $R>0$, and slack $\eta>0$ satisfying the parameter regime
\begin{equation}
\label{eq:lower-regime}
R\le k,
\qquad
\eta\le\frac{UR}{64k}.
\end{equation}
We define the coordinate resolution $\kappa$, alphabet size $M$, symbol centers $\mu_m$, and frame scale $\bar\theta$ by
\begin{align*}
\kappa := \frac{\eta}{U},
\qquad
M := \left\lfloor\frac{UR}{64k\eta}\right\rfloor,
\qquad
\mu_m := 4\kappa m\quad \forall m\in[M],
\qquad
\bar\theta := \frac{R}{16k}.
\end{align*}
Condition \eqref{eq:lower-regime} ensures $M\ge 1$. Moreover, consecutive symbol centers are separated by $4\kappa$, and every center lies in $(0,\bar\theta]$ since
\begin{align*}
\mu_M
&= 4\kappa M
\le 4\cdot\frac{\eta}{U}\cdot\frac{UR}{64k\eta}
= \frac{R}{16k}
= \bar\theta,
\end{align*}
where the inequality substitutes $\kappa=\eta/U$ and uses $M\le UR/(64k\eta)$.

\begin{definition}[Encoding frame]
\label{def:encoding-frame}
Let $K\subseteq\R^k$ be a closed convex cone. An \emph{encoding frame} for $K$ of size $n$ consists of a base point $x_0\in\R^k$ and orthonormal vectors $F_1,\ldots,F_n\in\R^k$, each orthogonal to $x_0$, such that for every $\theta\in[0,\bar\theta]^n$, the point
\begin{align*}
z_\theta := x_0 + \sum_{i=1}^n \theta_i F_i
\end{align*}
satisfies $z_\theta\in K$ and $\|z_\theta\|_2\le R$.
\end{definition}

For the nonnegative orthant $K=\R_+^k$, choosing $x_0=0$ and $F_i=e_i$ for $i\in[k]$ yields an encoding frame of size $n=k$: for every $\theta\in[0,\bar\theta]^k$, we have $z_\theta=\theta\in\R_+^k$ and
\begin{align*}
\|\theta\|_2
&\le \bar\theta\sqrt{k}
= \frac{R}{16\sqrt{k}}
\le R,
\end{align*}
where the first inequality bounds each coordinate by $\bar\theta$ and the last uses $k\ge 2$.

\paragraph{Hard instances.}
Fix an encoding frame of size $n$ and a multiplicity parameter $t\ge 1$. Given a hidden word $w=(w_1,\ldots,w_n)\in[M]^n$, the dataset $D_t(w)$ consists of $t$ identical copies, for each $i\in[n]$, of the paired constraints
\begin{align}
\langle UF_i,\,z\rangle &\le U\mu_{w_i},
\label{eq:lower-upper-row}\\
\langle -UF_i,\,z\rangle &\le -U\mu_{w_i},
\label{eq:lower-lower-row}
\end{align}
for a total of $2nt$ private rows. Each row $(a,b)$ satisfies $\|a\|_2=U$ because $\|F_i\|_2=1$, and its right-hand side satisfies
\begin{align*}
|b|
&= U\mu_{w_i}
\le U\bar\theta
= \frac{UR}{16k}
\le U,
\end{align*}
where the first inequality uses $\mu_{w_i}\le\bar\theta$ and the second uses $R\le k$ from \eqref{eq:lower-regime}.

To verify feasibility, set $z^\star(w):=z_\theta$ with $\theta_i:=\mu_{w_i}\in(0,\bar\theta]$. Because $F_1,\ldots,F_n$ are orthonormal and orthogonal to $x_0$, we have $\langle F_i,z^\star(w)\rangle=\mu_{w_i}$ for every $i\in[n]$, so $z^\star(w)$ satisfies all $2nt$ rows with equality. By \Cref{def:encoding-frame}, $z^\star(w)\in K$ and $\|z^\star(w)\|_2\le R$. Thus, every $D_t(w)$ is an $R$-feasible instance in the sense of \Cref{def:problem}. Note that scaling the rows to norm $U$ converts an additive row slack of $\eta$ into a coordinate tolerance of $\kappa=\eta/U$ along $F_i$, allowing the alphabet size $M$ to scale linearly with $U$, while the base point $x_0$ serves solely to keep the witnesses inside $K$.

\subsection{Decoding}
\label{subsec:lower-decoding}

Given any candidate point $\widehat z\in\R^k$, we define the \emph{decoded word} $q(\widehat z)\in[M]^n$ by rounding each frame projection to the nearest symbol center:
\begin{align*}
q_i(\widehat z)
&:= \arg\min_{m\in[M]} \bigl|\langle F_i,\widehat z\rangle - \mu_m\bigr|
\qquad (i\in[n]).
\end{align*}
We call coordinate $i\in[n]$ \emph{good} for $(\widehat z,w)$ if $\widehat z$ $\eta$-satisfies both \eqref{eq:lower-upper-row} and \eqref{eq:lower-lower-row} for coordinate $i$.

Whenever coordinate $i$ is good, $q_i(\widehat z)=w_i$ uniquely (with no ties). Indeed, dividing the two $\eta$-relaxed inequalities $\langle UF_i,\widehat z\rangle\le U\mu_{w_i}+\eta$ and $\langle -UF_i,\widehat z\rangle\le -U\mu_{w_i}+\eta$ by $U$ gives $|\langle F_i,\widehat z\rangle-\mu_{w_i}|\le\kappa$. Consequently, for every $m\in[M]\setminus\{w_i\}$,
\begin{align*}
\bigl|\langle F_i,\widehat z\rangle - \mu_m\bigr|
\ge |\mu_m - \mu_{w_i}| - \bigl|\langle F_i,\widehat z\rangle - \mu_{w_i}\bigr| \ge 4\kappa - \kappa
= 3\kappa
> \bigl|\langle F_i,\widehat z\rangle - \mu_{w_i}\bigr|,
\end{align*}
where the first inequality is the reverse triangle inequality, and the second uses that distinct centers are at least $4\kappa$ apart, while $|\langle F_i,\widehat z\rangle-\mu_{w_i}|\le\kappa$. Thus, $\mu_{w_i}$ is strictly closer to $\langle F_i,\widehat z\rangle$ than any other center.

\subsection{Dimension lower bound}
\label{subsec:lower-proof}

\begin{theorem}[Dimension lower bound]
\label{thm:lower-bound-main}
Assume \eqref{eq:lower-regime}, let $K\subseteq\R^k$ admit an encoding frame of size $n$, and fix $\varepsilon\in(0,1]$, $\delta\in[0,1]$, and $\beta_0\in(0,1)$. Define
\begin{align*}
t_\star
&:= \left\lfloor
\frac{1}{4\varepsilon}
\min\!\left\{
\log\!\frac{\beta_0 M}{4},\
\log\!\frac{\beta_0(e^\varepsilon-1)}{4\delta}
\right\}
\right\rfloor,
\end{align*}
interpreting the second argument of the minimum as $+\infty$ when $\delta=0$, and assume $t_\star\ge 1$. Then for every $(\varepsilon,\delta)$-DP algorithm $\mathcal M$ under add/remove row adjacency with outputs in $\R^k$, there exists $w\in[M]^n$ such that
\begin{align*}
\Pr\!\left[\operatorname{viol}_\eta\!\left(\mathcal M(D_{t_\star}(w));\,D_{t_\star}(w)\right)\ge\frac{nt_\star}{2}\right]
&> 1-\beta_0.
\end{align*}
\end{theorem}

Note that \Cref{thm:lower-bound-main} places no cone constraint on the output of $\mathcal M$, so it applies a fortiori to improper or proper solvers whose outputs lie in $K$.

\begin{proof}
Write $t=t_\star$ throughout. Suppose for contradiction that there exists an $(\varepsilon,\delta)$-DP algorithm $\mathcal M$ satisfying, for every $w\in[M]^n$,
\begin{align*}
\Pr\!\left[\operatorname{viol}_\eta\!\left(\mathcal M(D_t(w));\,D_t(w)\right) < \frac{nt}{2}\right]
&\ge \beta_0.
\end{align*}
Draw $W\sim\mathrm{Unif}([M]^n)$ and $I\sim\mathrm{Unif}([n])$ independently of each other and of the internal randomness of $\mathcal M$.

\emph{Correctness lower bound.} Because each distinct constraint in $D_t(W)$ is duplicated $t$ times, violating a constraint incurs all $t$ copies simultaneously. On the event 
\begin{align*}
    {\cal E}:=\left\{\operatorname{viol}_\eta(\mathcal M(D_t(W));D_t(W))<nt/2\right\},
\end{align*}
fewer than $n/2$ coordinates can have a violated constraint, leaving strictly more than $n/2$ good coordinates. By \Cref{subsec:lower-decoding}, $q_i(\mathcal M(D_t(W)))=W_i$ holds on every good coordinate $i$. Therefore,
\begin{align}
\Pr\left[q_I(\mathcal M(D_t(W)))=W_I\right]
\ge \Pr[{\cal E}]\cdot\Pr\left[q_I(\mathcal M(D_t(W)))=W_I \;\big|\; {\cal E}\right] 
> \beta_0 \cdot \frac{1}{2}
= \frac{\beta_0}{2},
\label{eq:lower-correctness}
\end{align}
where the strict inequality uses $\Pr[{\cal E}]\ge\beta_0$ by averaging the assumed success probability over $W$ and $\Pr[q_I(\mathcal M(D_t(W)))=W_I\mid {\cal E}]>1/2$ since $I$ is uniform on $[n]$ and independent of ${\cal E}$, while more than $n/2$ coordinates are decoded correctly on ${\cal E}$.

\emph{Group-privacy upper bound.} Let $S\sim\mathrm{Unif}([M])$ be independent of $(W,I,\mathcal M)$, and let $W'$ be obtained from $W$ by replacing $W_I$ with $S$. Since only the $2t$ rows associated with coordinate $I$ depend on $W_I$, transforming $D_t(W)$ into $D_t(W')$ requires deleting $2t$ rows and inserting $2t$ rows, so their add/remove distance is at most $4t$. Conditioning on $I$ and applying group privacy (\Cref{fact:group-privacy}) to the event $\{q_I(\mathcal M(\cdot))=W_I\}$ yields
\begin{align}
\Pr\bigl[q_I(\mathcal M(D_t(W)))=W_I\bigr]
&\le e^{4t\varepsilon}\,\Pr\bigl[q_I(\mathcal M(D_t(W')))=W_I\bigr] + \delta_{4t} \nonumber\\
&= e^{4t\varepsilon}\cdot\frac{1}{M} + \frac{\delta\,(e^{4t\varepsilon}-1)}{e^\varepsilon-1} \nonumber\\
&\le e^{4t\varepsilon}\!\left(\frac{1}{M}+\frac{\delta}{e^\varepsilon-1}\right),
\label{eq:lower-group-ineq}
\end{align}
where the first step applies group privacy at distance $4t$ with $\delta_{4t}=\delta(e^{4t\varepsilon}-1)/(e^\varepsilon-1)$ which equals $0$ when $\delta=0$, the equality uses that $D_t(W')$ is independent of $W_I\sim\mathrm{Unif}([M])$, so any output matches $W_I$ with probability exactly $1/M$, and the final inequality follows from $e^{4t\varepsilon}-1\le e^{4t\varepsilon}$.

\emph{Contradiction.} Chaining \eqref{eq:lower-correctness} and \eqref{eq:lower-group-ineq} gives
\begin{align*}
\frac{\beta_0}{2}
&< e^{4t\varepsilon}\!\left(\frac{1}{M}+\frac{\delta}{e^\varepsilon-1}\right)
\le \frac{\beta_0}{4} + \frac{\beta_0}{4}
= \frac{\beta_0}{2},
\end{align*}
where the second inequality follows from the definition of $t=t_\star$, which ensures both $e^{4t\varepsilon}/M\le\beta_0/4$ and $e^{4t\varepsilon}\delta/(e^\varepsilon-1)\le\beta_0/4$ (the latter holding trivially when $\delta=0$). This contradiction completes the proof of the theorem.
\end{proof}

\begin{corollary}[Asymptotic form]
\label{cor:lower-add-remove}
Assume \eqref{eq:lower-regime}, let $K\subseteq\R^k$ admit an encoding frame of size $n$, let $\varepsilon\in(0,1]$, and suppose further that
\begin{align*}
\frac{UR}{k\eta} \ge 2^{20}
\qquad\text{and}\qquad
\frac{\varepsilon}{\delta} \ge 2^{9}.
\end{align*}
Let $t_\star$ be defined as in \Cref{thm:lower-bound-main} with $\beta_0=1/2$. If an $(\varepsilon,\delta)$-DP algorithm $\mathcal M$ under add/remove row adjacency satisfies
\begin{align*}
\Pr\bigl[\operatorname{viol}_\eta(\mathcal M(D);D)\le s\bigr]
&\ge \frac{1}{2}
\qquad\text{for all }D\in\bigl\{D_{t_\star}(w):w\in[M]^n\bigr\},
\end{align*}
then
\begin{align*}
s
&\ge \frac{n}{32\,\varepsilon}\,\log\min\!\left\{\frac{UR}{k\eta},\ \frac{\varepsilon}{\delta}\right\}.
\end{align*}
\end{corollary}

\begin{proof}
We apply \Cref{thm:lower-bound-main} with $\beta_0=1/2$. It suffices to verify $t_\star\ge 1$ and lower-bound $t_\star$. Let $A:=UR/(k\eta)\ge 2^{20}$ and $B:=\varepsilon/\delta\ge 2^9$. For $\beta_0=1/2$, the two arguments inside the minimum defining $t_\star$ are $\log(M/8)$ and $\log\bigl((e^\varepsilon-1)/(8\delta)\bigr)$, which we bound in turn:
\begin{itemize}
\item For the first term,
\begin{align*}
\log\frac{M}{8}
&\ge \log\frac{A}{1024}
= \log A - 10\log 2
\ge \max\!\left\{\frac{1}{2}\log A,\, 10\log 2\right\},
\end{align*}
where the first inequality uses $M=\lfloor A/64\rfloor\ge A/128$ (since $A/64\ge 2^{14}$), and the second uses $\log A\ge 20\log 2$. Note that $10\log 2 > 4$.

\item For the second term,
\begin{align*}
\log\frac{e^\varepsilon-1}{8\delta}
&\ge \log\frac{\varepsilon}{8\delta}
= \log B - 3\log 2
\ge \max\!\left\{\frac{1}{2}\log B,\, 6\log 2\right\},
\end{align*}
where the first inequality uses $e^\varepsilon-1\ge\varepsilon$, and the second uses $\log B\ge 9\log 2$ (so $3\log 2\le\frac{1}{3}\log B\le\frac{1}{2}\log B$ and $\log B - 3\log 2\ge 6\log 2$). Again, $6\log 2 > 4$.
\end{itemize}
Combining both bounds, the quantity inside the floor defining $t_\star$ is at least $4/(4\varepsilon)=1/\varepsilon\ge 1$ (since $\varepsilon\le 1$) and at least $\frac{1}{8\varepsilon}\log\min\{A,B\}$. Applying $\lfloor x\rfloor\ge x/2$ for $x\ge 1$ yields $t_\star\ge 1$ and
\begin{align*}
s
&\ge \frac{nt_\star}{2}
\ge \frac{n}{2}\cdot\frac{1}{16\varepsilon}\log\min\{A,B\}
= \frac{n}{32\varepsilon}\log\min\{A,B\},
\end{align*}
where the first inequality follows from \Cref{thm:lower-bound-main} since $\Pr[\operatorname{viol}_\eta\le s]\ge 1/2 = 1-\beta_0$.
\end{proof}

Instantiating \Cref{cor:lower-add-remove} with the orthant frame ($n=k$) establishes \Cref{thm:lower-informal}, where the hard instances are linear programs (i.e., SCPs over $\R_+^k$). When both $UR/(k\eta)$ and $\varepsilon/\delta$ are bounded constants at the edge of the regime, the lower bound reduces to $\Omega(k/\varepsilon)$, and decreasing $\eta$ and $\delta$ together strengthens it logarithmically.

\subsection{Extension to non-polyhedral public cones}
\label{subsec:lower-nonpolyhedral}

Because \Cref{thm:lower-bound-main} and \Cref{cor:lower-add-remove} depend on the public cone $K$ solely through the size $n$ of its encoding frame, constructing frames of size $n=k-1$ immediately transfers the lower bound (up to constant factors) to single Lorentz and PSD cones.

\begin{corollary}[Lower bound for single Lorentz and PSD cones]
\label{cor:lower-lorentz-psd}
\Cref{thm:lower-bound-main} and \Cref{cor:lower-add-remove} hold with $n=k-1$ when the public cone is either:
\begin{enumerate}[leftmargin=2em,topsep=2pt,itemsep=2pt]
\item[(a)] the Lorentz cone $\mathcal L^k:=\{(s,y)\in\R\times\R^{k-1}:s\ge\|y\|_2\}$ (rank $2$), or
\item[(b)] the PSD cone $\mathbb S_+^r$ equipped with the trace inner product, where $k=r(r+1)/2$ (rank $r$).
\end{enumerate}
\end{corollary}

\begin{proof}
In each case, we construct an encoding frame (\Cref{def:encoding-frame}) of size $n=k-1$, recalling $\bar\theta=R/(16k)$.

\emph{(a) Lorentz cone.} Under the coordinates of \Cref{subsec:eja-background}, the Jordan trace norm coincides with the standard Euclidean norm on $\R^k$. Choose $x_0=(R/\sqrt{2},\,0)$ and $F_i=(0,e_i)$ for $i\in[k-1]$, which are orthonormal and orthogonal to $x_0$. For any $\theta\in[0,\bar\theta]^{k-1}$, writing $z_\theta=(R/\sqrt{2},\,\theta)$, we have
\begin{align*}
\|\theta\|_2
&\le \bar\theta\sqrt{k}
= \frac{R}{16\sqrt{k}}
\le \frac{R}{\sqrt{2}},
\\
\|z_\theta\|_2^2
&= \frac{R^2}{2} + \|\theta\|_2^2
\le \frac{R^2}{2} + \frac{R^2}{2}
= R^2,
\end{align*}
where the first line bounds each of the $k-1<k$ coordinates by $\bar\theta$ and uses $16\sqrt{k}\ge\sqrt{2}$. The first bound shows $z_\theta\in\mathcal L^k$, and the second gives $\|z_\theta\|_2\le R$.

\emph{(b) PSD cone.} Choose $X_0=\frac{R}{\sqrt{2r}}\,I_r$, which satisfies $\|X_0\|_F=R/\sqrt{2}$ and $\lambda_{\min}(X_0)=R/\sqrt{2r}$, and let $F_1,\ldots,F_{k-1}$ be a Frobenius-orthonormal basis of the traceless symmetric subspace $\{X\in\mathbb S^r:\mathrm{tr}[X]=0\}=I_r^\perp$. For any $\theta\in[0,\bar\theta]^{k-1}$, let $\Delta:=\sum_{i=1}^{k-1}\theta_i F_i$. Then
\begin{align*}
\|\Delta\|_{\mathrm{op}}
&\le \|\Delta\|_F
\le \bar\theta\sqrt{k}
= \frac{R}{16\sqrt{k}}
\le \frac{R}{\sqrt{2r}},
\\
\lambda_{\min}(X_0+\Delta)
&\ge \lambda_{\min}(X_0) - \|\Delta\|_{\mathrm{op}}
\ge 0,
\\
\|X_0+\Delta\|_F^2
&= \|X_0\|_F^2 + \|\Delta\|_F^2
\le \frac{R^2}{2} + \frac{R^2}{2}
= R^2.
\end{align*}
Here the first line uses $\|\Delta\|_{\mathrm{op}}\le\|\Delta\|_F$, orthonormality of $\{F_i\}$ with $k-1<k$, and $k=r(r+1)/2\ge r$; the second line applies Weyl's inequality together with $\lambda_{\min}(X_0)=R/\sqrt{2r}$ to conclude $z_\theta=X_0+\Delta\succeq 0$; and the third line uses Frobenius orthogonality ($\langle X_0,\Delta\rangle\propto\mathrm{tr}[\Delta]=0$) to conclude $\|z_\theta\|_F\le R$.
\end{proof}

Note that for $\mathbb S_+^r$, embedding an orthant along the diagonal would yield only $\Omega(r/\varepsilon)$ violations, whereas using the full traceless subspace $I_r^\perp$ packs the off-diagonal entries as well to give $\Omega(k/\varepsilon)=\Omega(r^2/\varepsilon)$.

\section{A Complexity-Theoretic Barrier to Stronger Lower Bounds}
\label{app:np-barrier}

The exponential mechanism in \Cref{rmk:lower-tight} achieves strong accuracy under pure differential privacy, but sampling from it directly takes time exponential in $k$. In this appendix, we show that for linear programs, a discretized version of this mechanism with weights proportional to $2^{-\viol_\eta}$ reduces to uniform generation of $\NP$ witnesses, and therefore runs in randomized polynomial time if $\NP\subseteq\BPP$ (\Cref{prop:np-bpp}). Consequently, any unconditional proof that polynomial-time $(1,\delta)$-DP algorithms must incur $\omega\bigl(k\log(2+kUR/\eta)\bigr)$ violations would imply $\NP\not\subseteq\BPP$, and hence $\Ptime\ne\NP$ (\Cref{cor:np-barrier}).

Throughout this appendix, the public cone is $K=\R_+^k$ and all input numbers are rationals encoded in binary. An input consists of the dimension $k$ in unary, the public parameters $U,R,\eta>0$, and the private rows $(a_i,b_i)_{i\in[m]}$ satisfying $\|a_i\|_2,|b_i|\le U$. Let $L$ denote the total bit length of the input. As in the rest of the paper, only the output point $\widehat x$ is released. 


\begin{proposition}[Efficient private LP feasibility under $\NP\subseteq\BPP$]
\label{prop:np-bpp}
Assume $\NP\subseteq\BPP$, and let
\begin{align*}
M:=\left\lceil\frac{2kUR}{\eta}\right\rceil
\qquad\text{and}\qquad
\ell:=k\,\bigl\lceil\log_2(M+1)\bigr\rceil.
\end{align*}
There is a polynomial-time algorithm that is $(1,0)$-DP under add/remove row adjacency on all inputs and always outputs a point $\widehat x\in\R_+^k$ with $\|\widehat x\|_2\le R$. Moreover, on every $R$-feasible instance (\Cref{def:problem}) and for every $\beta\in(0,1)$, with probability at least $1-\beta$,
\begin{align*}
\viol_\eta(\widehat x)\;\le\;\ell+\log_2\frac{2}{\beta}
\;=\;O\!\left(k\log\Bigl(2+\frac{kUR}{\eta}\Bigr)+\log\frac{1}{\beta}\right).
\end{align*}
\end{proposition}

\begin{corollary}[A barrier to stronger lower bounds]
\label{cor:np-barrier}
Any unconditional proof that, for some $\delta\ge 0$, no polynomial-time $(1,\delta)$-DP algorithm satisfies $\Pr[\viol_\eta(\widehat x)\le\ell+3]\ge 2/3$ on every $R$-feasible LP instance would imply $\NP\not\subseteq\BPP$, and in particular $\Ptime\ne\NP$. 

In particular, when $U=R=1$ and $\eta=k^{-2}$, we have $\ell=O(k\log k)$, so proving an unconditional $\omega(k\log k)$ violation lower bound for polynomial-time $(1,\delta)$-DP algorithms in this regime would prove $\Ptime\ne\NP$.
\end{corollary}

\begin{proof}
Under $\NP\subseteq\BPP$, running the algorithm of \Cref{prop:np-bpp} with $\beta=1/3$ yields $\viol_\eta(\widehat x)\le\ell+\log_2 6\le\ell+3$ with probability at least $2/3$, while satisfying $(1,\delta)$-DP for every $\delta\ge 0$. Any unconditional lower bound ruling out such an algorithm therefore implies $\NP\not\subseteq\BPP$, and since $\Ptime\subseteq\BPP$, also $\Ptime\ne\NP$. When $U=R=1$ and $\eta=k^{-2}$, we have $M=2k^3$ and hence $\ell=k\lceil\log_2(2k^3+1)\rceil=O(k\log k)$.
\end{proof}


\subsection{An ideal mechanism on a grid}
\label{subsec:np-grid}

Consider the discretized domain
\[
\mathcal G:=\left\{\frac{R}{M}\,z:\ z\in\mathbb Z_{\ge 0}^k,\ \|z\|_2\le M\right\}\;\subseteq\;\R_+^k\cap B_R(0).
\]
Each coordinate $z_j$ belongs to $\{0,\ldots,M\}$ and can be represented using $\lceil\log_2(M+1)\rceil$ bits, which gives every point of $\mathcal G$ a canonical $\ell$-bit encoding and also implies $|\mathcal G|\le 2^\ell$. %
Membership in $\mathcal G$ reduces to an exact integer norm test, and although $M$ may be exponential in $L$, the encoding length $\ell$ remains polynomial in $L$.

\begin{lemma}[Rounding to the grid]
\label{lem:np-grid}
If $D$ is $R$-feasible, then there exists $g\in\mathcal G$ with $\viol_\eta(g;D)=0$.
\end{lemma}

\begin{proof}
Let $x^\star\in\R_+^k$ witness $R$-feasibility of $D$, and define $g:=(R/M)z$ by coordinatewise rounding down, $z_j:=\lfloor Mx^\star_j/R\rfloor$. Then $0\le g\le x^\star$ coordinatewise, so $\|z\|_2\le M\|x^\star\|_2/R\le M$ and thus $g\in\mathcal G$. Since $|g_j-x^\star_j|<R/M$ for each $j\in[k]$, we also have $\|g-x^\star\|_2<R\sqrt{k}/M$. Hence, for every row $i\in[m]$,
\begin{align*}
\langle a_i,g\rangle\le\langle a_i,x^\star\rangle+U\|g-x^\star\|_2
<b_i+\frac{UR\sqrt{k}}{M}
\le b_i+\frac{\eta}{2\sqrt{k}}
<b_i+\eta. 
\end{align*}
Thus, $\viol_\eta(g;D)=0$ and the lemma is proved.
\end{proof}

For any dataset $D$ (feasible or not), define the base-$2$ exponential mechanism distribution over $\mathcal G$ by
\[
\pi_D(x):=\frac{2^{-\viol_\eta(x;D)}}{Z_D},
\qquad
Z_D:=\sum_{y\in\mathcal G}2^{-\viol_\eta(y;D)}.
\]

\begin{lemma}[Properties of $\pi_D$]
\label{lem:np-ideal}
Let $D$ be any dataset with $m$ rows.
\begin{enumerate}[label=(\roman*)]
\item If $D'$ differs from $D$ by adding or removing one row, then $\frac{1}{2}\le\pi_{D'}(x)/\pi_D(x)\le 2$ for all $x\in\mathcal G$.
\item $\pi_D(x)\ge 2^{-(m+\ell)}$ for all $x\in\mathcal G$.
\item If $D$ is $R$-feasible, then $\Pr_{x\sim\pi_D}[\viol_\eta(x;D)\ge j]\le 2^{\ell-j}$ for every integer $j\ge 0$.
\end{enumerate}
\end{lemma}

\begin{proof}
(i) By symmetry, assume $D'$ is obtained by adding one row to $D$. Each unnormalized weight $2^{-\viol_\eta(x;\cdot)}$ is multiplied by either $1$ or $\frac{1}{2}$, so $Z_{D'}/Z_D\in[\frac{1}{2},1]$ and the ratio $\pi_{D'}(x)/\pi_D(x)$ lies in $[\frac{1}{2},2]$.

(ii) Every unnormalized weight is at least $2^{-m}$, while $Z_D\le|\mathcal G|\le 2^\ell$.

(iii) By \Cref{lem:np-grid}, $Z_D\ge 1$, whereas each of the at most $|\mathcal G|\le 2^\ell$ points with $\viol_\eta(x;D)\ge j$ contributes weight at most $2^{-j}$.
\end{proof}

\subsection{Sampling by uniform generation of witnesses}
\label{subsec:np-sampling}

Fix a dataset $D$ with $m$ rows. Call a pair $(c,\omega)\in\{0,1\}^\ell\times\{0,1\}^m$ \emph{accepted} if $c$ encodes a grid point $x\in\mathcal G$ and $\omega_i=0$ for every row $i\in[m]$ violated by $x$ (i.e., $\langle a_i,x\rangle>b_i+\eta$). Acceptance can be verified in polynomial time using exact rational arithmetic, and each pair has length $\ell+m\le\mathrm{poly}(L)$, so this defines an $\NP$ relation. Note that the witness set is always nonempty, since the encoding of $x=0$ paired with $\omega=0^m$ is always accepted. For a fixed $x\in\mathcal G$, the bits $\omega_i$ on non-violated rows are unconstrained, so $x$ admits exactly $2^{m-\viol_\eta(x;D)}$ accepted pairs. Thus, drawing $(c,\omega)$ uniformly at random from all accepted pairs and returning the grid point encoded by $c$ samples exactly from $\pi_D$.

\begin{fact}[Uniform generation with an $\NP$ oracle \cite{bellare2000uniform}]
\label{fact:bgp}
For every $\NP$ relation, there is a randomized polynomial-time algorithm with a SAT oracle that, on any input having at least one witness, outputs either a witness or a failure symbol $\bot$. Its failure probability is bounded above by a constant strictly less than $1$, and conditioned on not failing, its output is distributed uniformly over all witnesses.
\end{fact}

\paragraph{The algorithm.}
Set $\nu_D:=2^{-(m+\ell+4)}$. Given an exact SAT oracle, we run the generator of \Cref{fact:bgp} on the relation above for $O(m+\ell)$ independent trials, choosing the implied constant so that all trials fail simultaneously with probability at most $\nu_D/2$, and return the grid point of the first accepted pair (or $0$ if all trials fail). Since the output equals $\pi_D$ conditioned on at least one trial succeeding, this exact-oracle procedure has output distribution within total variation distance $\nu_D/2$ of $\pi_D$.

Under $\NP\subseteq\BPP$, we replace each SAT oracle query with the majority vote of $O(\log(N/\nu_D))$ independent runs of a $\BPP$ decider for SAT, where $N=\mathrm{poly}(L)$ bounds the total number of oracle queries across all trials. By a Chernoff bound, each simulated answer errs with probability at most $\nu_D/(2N)$ on any given query. We also enforce the runtime budget of the exact-oracle procedure, returning $0$ on timeout or if the resulting pair fails verification. Because $\log_2(1/\nu_D)=m+\ell+4\le\mathrm{poly}(L)$, this simulated procedure always terminates in polynomial time and outputs a point in $\mathcal G$. Coupling the simulated and exact-oracle executions under shared generator randomness, the two runs agree unless at least one simulated oracle answer errs; even with adaptively chosen queries, a union bound over the at most $N$ queries limits this event to probability at most $\nu_D/2$. By the triangle inequality, the output distribution $\widehat\pi_D$ of the algorithm satisfies
\[
\operatorname{TV}(\widehat\pi_D,\pi_D)\;\le\;\nu_D.
\]

\subsection{\texorpdfstring{Proof of \Cref{prop:np-bpp}}{Proof of Proposition E.1}}
\label{subsec:np-proof}
\begin{proof}
By construction, the algorithm runs in polynomial time and always outputs $\widehat x\in\mathcal G\subseteq\R_+^k\cap B_R(0)$.

\emph{Privacy.} Combining $\operatorname{TV}(\widehat\pi_D,\pi_D)\le\nu_D$ with \Cref{lem:np-ideal}(ii) gives $|\widehat\pi_D(x)-\pi_D(x)|\le\nu_D\le\pi_D(x)/16$ for every $x\in\mathcal G$, so
\[
\frac{15}{16}\,\pi_D(x)\;\le\;\widehat\pi_D(x)\;\le\;\frac{17}{16}\,\pi_D(x)
\qquad\text{for all }x\in\mathcal G.
\]
For any add/remove neighbors $D$ and $D'$, applying \Cref{lem:np-ideal}(i) yields
\[
\widehat\pi_D(x)\;\le\;\frac{17}{16}\,\pi_D(x)\;\le\;\frac{17}{16}\cdot 2\,\pi_{D'}(x)\;\le\;\frac{34}{15}\,\widehat\pi_{D'}(x)
\qquad\text{for all }x\in\mathcal G.
\]
Summing over any subset of $\mathcal G$ establishes $(\ln\frac{34}{15},0)$-DP, and hence $(1,0)$-DP since $\ln\frac{34}{15}<1$. 

\emph{Utility.} For an $R$-feasible instance $D$, \Cref{lem:np-ideal}(iii) and the upper bound above imply 
\begin{align*}
    \Pr_{\widehat x\sim\widehat\pi_D}\left[\viol_\eta(\widehat x;D)\ge j\right]\le\frac{17}{16}\,2^{\ell-j}\qquad \forall j\geq 0.
\end{align*}
Choosing $j:=\ell+\lfloor\log_2(2/\beta)\rfloor+1$ ensures $2^{\ell-j}<\beta/2$, so this probability is at most $\frac{17}{32}\,\beta<\beta$. Since $\viol_\eta(\widehat x;D)$ is integer-valued, we obtain $\viol_\eta(\widehat x;D)\le j-1\le\ell+\log_2(2/\beta)$ with probability at least $1-\beta$. Finally, $M+1\le 2+2kUR/\eta$ gives
\begin{align*}
    \ell\le k\bigl(1+\log_2(2+2kUR/\eta)\bigr)=O\bigl(k\log(2+kUR/\eta)\bigr).
\end{align*}

Therefore, the proof of the proposition is completed.
\end{proof}
\end{document}